\documentclass[11pt,a4paper]{article}
\usepackage{jinstpub}
\usepackage{siunitx}
\usepackage{subcaption} 
\usepackage{hyperref}
\usepackage{upgreek}
\usepackage{placeins}
\usepackage{floatrow}

\newfloatcommand{capbtabbox}{table}[][\FBwidth]

\newif\ifproofread
\newcommand{\changemarker}[1]{%
\ifproofread
\textcolor{red}{#1}%
\else
#1%
\fi
}

\usepackage{lineno}
\title{The CGEM-IT readout chain}
\date{May 2021}
\author[a,b]{A.~Amoroso}
\author[c]{, R.~Baldini~Ferroli }
\author[d,e]{, I.~Balossino }
\author[c]{, M.~Bertani }
\author[d]{, D.~Bettoni}
\author[a,b]{, F.~Bianchi}
\author[a,b,1]{A.~Bortone\note{Corresponding author.}}
\author[i]{, R.~Bugalho}
\author[c]{, A.~Calcaterra}
\author[c]{, S.~Cerioni\footnote{Deceased}}
\author[c]{, S.~Chiozzi}
\author[d]{, G.~Cibinetto}
\author[d]{, A.~Cotta~Ramusino}
\author[b]{, F.~Cossio}
\author[b]{, M.~Da~Rocha~Rolo}
\author[a,b]{, F.~De~Mori}
\author[a,b]{, M.~Destefanis}
\author[i]{, A.~Di~Francesco}
\author[d,f]{, F.~Evangelisti}
\author[d,f]{, R.~Farinelli}
\author[b,j]{, L.~Fava}
\author[c]{, G.~Felici~ }
\author[b]{, S.~Garbolino}
\author[d,f]{, I.~Garzia}
\author[c]{, M.~Gatta}
\author[b]{, G.~Giraudo}
\author[d,f]{, S.~Gramigna}
\author[a,b]{, M.~Greco}
\author[a,b]{, L.~Lavezzi}
\author[a,b]{, M.~Maggiora}
\author[d]{, R.~Malaguti}
\author[g,h]{, A.~Mangoni}
\author[a,b]{, S.~Marcello}
\author[k]{, P.~Marciniewski}
\author[d]{, M.~Melchiorri}
\author[d,e]{, G.~Mezzadri}
\author[a,b]{, M.~Mignone}
\author[a,b]{, S.~Morgante}
\author[c]{, E.~Pace~ }
\author[g,h]{, S.~Pacetti}
\author[c]{, P.~Patteri}
\author[b]{, A.~Rivetti}
\author[d,f]{, M.~Scodeggio}
\author[a,b]{, S.~Sosio}
\author[a,b]{, S.~Spataro}
\author[i]{, J.~Varela}
\author[b]{, R.~Wheadon}

\affiliation[a]{Università  di Torino, Dipartimento di Fisica,\\ via P. Giuria 1, 10125 Torino, Italy
}

\affiliation[b]{INFN, Sezione di Torino, \\via P. Giuria 1, 10125 Torino, Italy}
\affiliation[c]{INFN, Laboratori Nazionali di Frascati,\\ via E. Fermi 40, 00044 Frascati (Roma), Italy
}
\affiliation[d]{INFN, Sezione di Ferrara,\\ via G. Saragat 1, 44122 Ferrara, Italy
}

\affiliation[e]{Institute of High Energy Physics, Chinese Academy of Sciences,\\ 19B YuquanLu, Beijing, 100049, People’s Republic of China
}

\affiliation[f]{Università di Ferrara, Dipartimento di Fisica e Scienze della Terra,\\ via G. Saragat 1, 44122 Ferrara, Italy
}
\affiliation[g]{INFN, Sezione di Perugia,\\ via A. Pascoli, 06123 Perugia, Italy
}
\affiliation[h]{Università di Perugia, Dipartimento di Fisica e Geologia,\\ via A. Pascoli, 06123 Perugia, Italy
}

\affiliation[i]{Laboratório de Instrumentação e Física Experimental de Partículas (LIP),\\ Av Elias Garcia 14, 1000-149,
Portugal

}
\affiliation[j]{Università del Piemonte Orientale, Dipartimento di Scienze e Innovazione Tecnologica, Viale Teresa Michel 11, 15121 Alessandria, Italy}

\affiliation[k]{Department of Physics and Astronomy, Uppsala University, \\ Lägerhyddsvägen 1, 752 37 Uppsala, Sweden}

\emailAdd{abortone@to.infn.it}

\abstract{An innovative Cylindrical Gas Electron Multiplier (CGEM) detector is under construction for the upgrade of the inner tracker of the BESIII experiment. A novel system has been worked out for the readout of the CGEM detector, including a new ASIC, dubbed TIGER -Torino Integrated GEM Electronics for Readout, designed for the amplification and digitization of the CGEM output signals. The data output by TIGER are collected and processed by a first FPGA-based module, GEM Read Out Card, in charge of configuration and control of the front-end ASICs. A second FPGA-based module, named GEM Data Concentrator, builds the trigger selected event packets containing the data and stores them via the main BESIII data acquisition system. The design of the electronics chain, including  the power and signal distribution, will be presented together with its performance.}
\begin{document}
\proofreadfalse
\maketitle
\flushbottom
\section{Introduction}
\begin{figure}
\centering
 \includegraphics[width=0.6\linewidth]{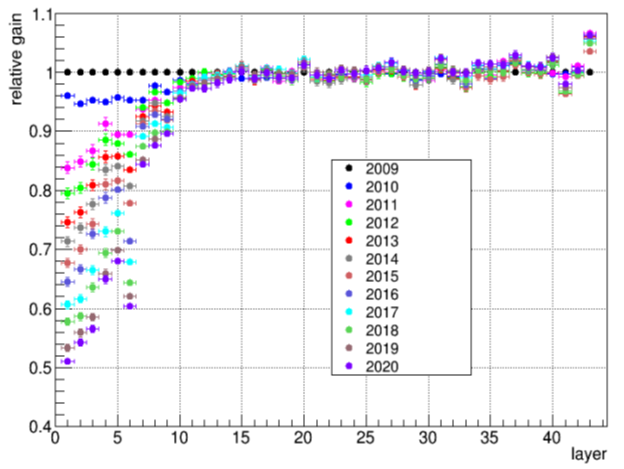}
  \caption{Gain loss of the MDC with respect to 2009. The inner MDC is composed of the eight innermost layers.  \changemarker{Their gain loss per year is about \SI{4}{\percent}}.}
\label{fig:gain_loss}

\end{figure}

\begin{figure}
\begin{floatrow}
\ffigbox{%
  \includegraphics[width=1\linewidth]{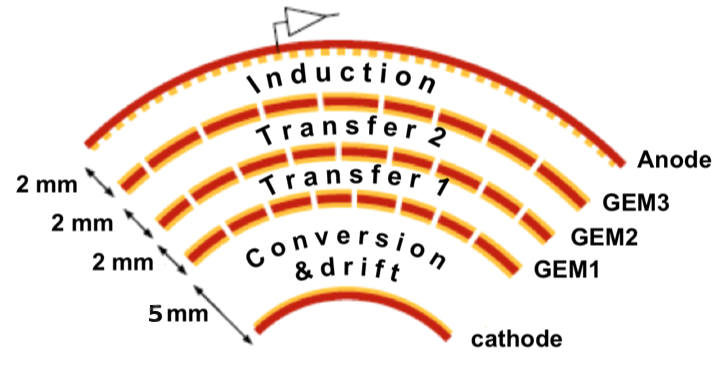}
  \caption{\changemarker{Single layer triple-GEM layout. Each detector layer is composed by cathode, anode and 3 GEM multiplication foils.}}
 \label{fig:CGEM_spac}%
}

\capbtabbox{%
  \begin{tabular}{cc|cc}
\multicolumn{2}{c|}{GEM voltages} & \multicolumn{2}{c}{Fields} \\ \hline               &                & Induction           & \SI{5}{\kilo \volt \per \centi \meter}       \\ \hline
G3             & \SI{275}{\volt}           &                 &          \\ \hline
               &                & Transfer 2      & \SI{3}{\kilo \volt \per \centi \meter}       \\ \hline
G2             & \SI{280}{\volt}              &                 &          \\ \hline
               &                & Transfer 1      & \SI{3}{\kilo \volt \per \centi \meter}        \\ \hline
G1             & \SI{280}{\volt}              &                 &          \\ \hline
               &                & Drift       & \SI{1.5}{\kilo \volt \per \centi \meter}     
\end{tabular}
}{%
\caption{ \changemarker{Voltage across the GEM foils and transfer fields. The total GEM voltage of \SI{835}{\volt} corresponds to a gain around 13500.}}
    \label{table:cgem_field}
    }
\end{floatrow}
\end{figure}

\begin{figure}
\centering
\begin{subfigure}{.4\textwidth}
  \centering
    \includegraphics[width=0.9\linewidth]{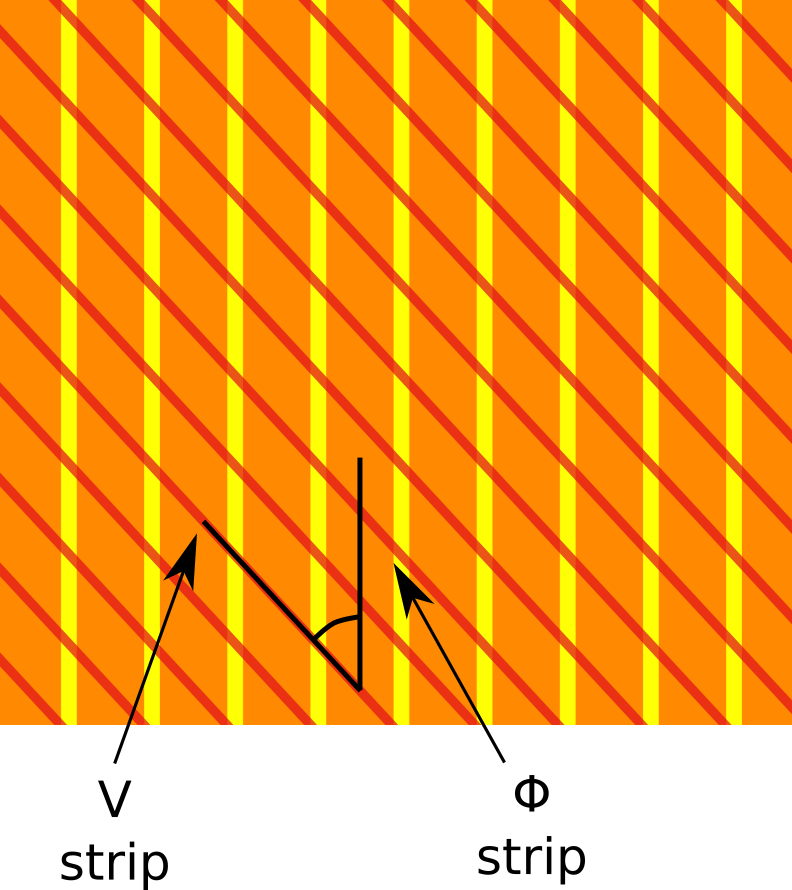}
   \caption{}
   \label{fig:strips_square}
   \end{subfigure}%
    \begin{subfigure}{.6\textwidth}
  \centering
    \includegraphics[width=0.9\linewidth]{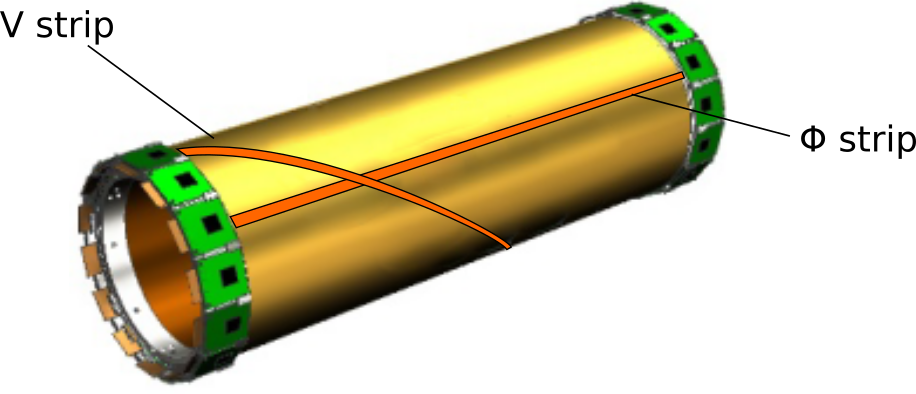}
      \caption{}
       \label{fig:strips_GEM}
        \end{subfigure}%
  \caption{\changemarker{Strip layout. (a): the anode structure for layer 1, with the $\Upphi$ strips in orange, the V strips in red and the \SI{46.7}{\degree} angle between them in black; (b): a sketch of the anode strips orientation on the cylindrical structure.}}
 \label{fig:strips}
\end{figure}

The Beijing Spectrometer (BESIII) is a high-energy physics experiment located at the Beijing Electron Positron Collider (BEPCII) \cite{bepcII2009}. 
BEPCII is a two-ring $e^+$ $e^-$ collider, with a center of mass energy tunable between 2 and \SI{4.9}{\giga \electronvolt} and a luminosity up to \SI{e33}{\cm^{-2}  \second ^{-1}}.
The cylindrical core of the BESIII detector consists of a helium-based  \changemarker{(He-C$_3$H$_8$ 60:40)} multilayer drift chamber (MDC), as charged particle tracker, a plastic scintillator time-of-flight system, and a CsI(Tl) electromagnetic calorimeter (EMC), which are all enclosed in a superconducting solenoidal magnet providing a \SI{1}{\tesla} magnetic field. The solenoid is supported by an octagonal flux-return yoke with resistive plate counter muon identifier modules interleaved with steel.  The acceptance of charged particles and  \changemarker{gammas} is \SI{93}{\percent} over the 4$\uppi$ solid angle.
The charged-particle momentum resolution at \SI{1}{\giga \electronvolt \per c} is \SI{0.5}{\percent}, and  the dE/dx resolution is \SI{6}{\percent} for the electrons from Bhabha scattering. More details about the BESIII spectrometer are described in \cite{Ablikim2010}.
\\
The BESIII experiment offers a unique setup to investigate charmonium and charmonium-like states, charmed mesons and baryons, light hadron spectroscopy, $\tau$ physics, QCD and CKM parameters, baryon form factors, and new physics by studying rare and forbidden decays \cite{White_paper}. \\
The current tracker, consisting of an inner chamber and an outer chamber, has been taking data since 2009. Due to the high luminosity of the experiment, it is degrading with a gain loss per year of about \SI{4}{\percent} for the innermost layers, as shown in figure \ref{fig:gain_loss} \cite{Dong_2016}. To partially compensate the gain loss, the voltage has been raised and about 0.2\%  water  vapor was added  to the MDC  gas  mixture  to  solve  the  cathode  aging problem.\\
Since BESIII is expected to be in operation for the next ten years, an upgrade is needed. An innovative solution to replace the aging inner MDC is to  use a CGEM Inner Tracker (CGEM-IT). It was proposed by the Italian Collaboration in BESIII and boosted by a European-Chinese network, funded by the European Commission. Indeed, GEM technology allows larger rate capability and reduced aging effects on the long period, therefore a longer lifespan.\\
\begin{figure}
\centering
    \begin{subfigure}{.6\textwidth}
  \centering
    \includegraphics[width=0.9\linewidth]{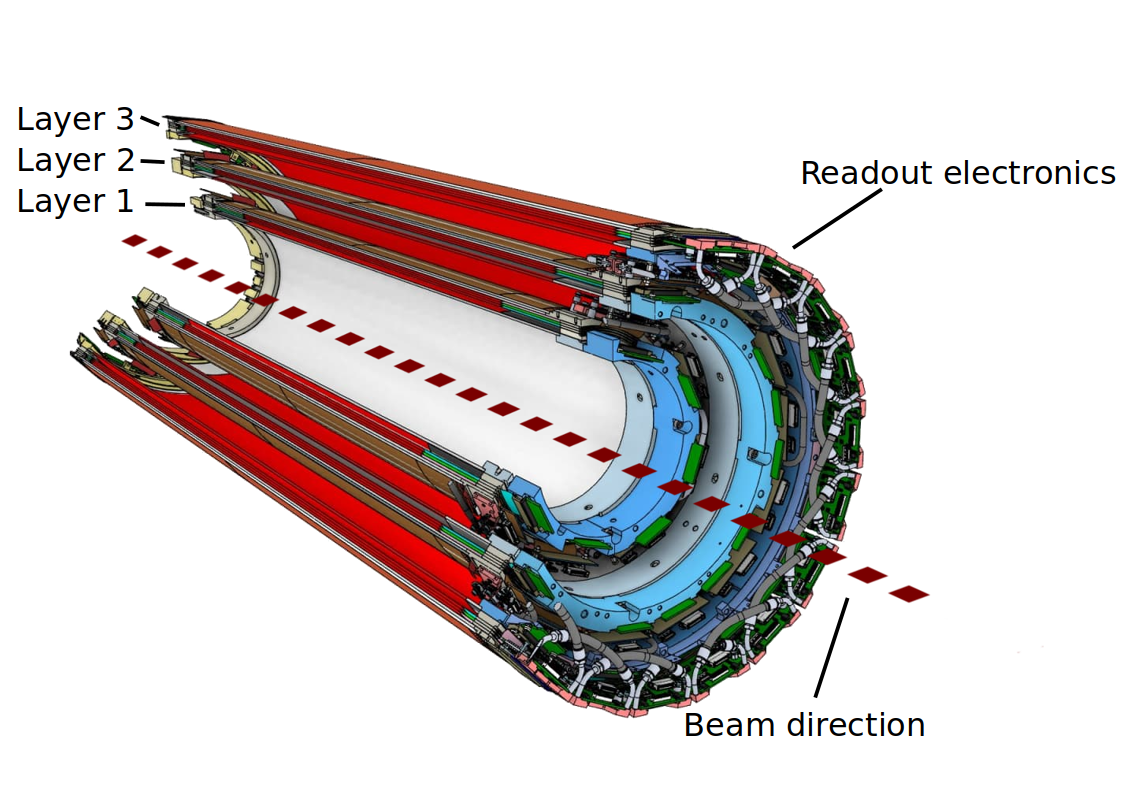}
        \end{subfigure}%
 \begin{subfigure}{.4\textwidth}
  \centering
      \includegraphics[width=0.8\linewidth]{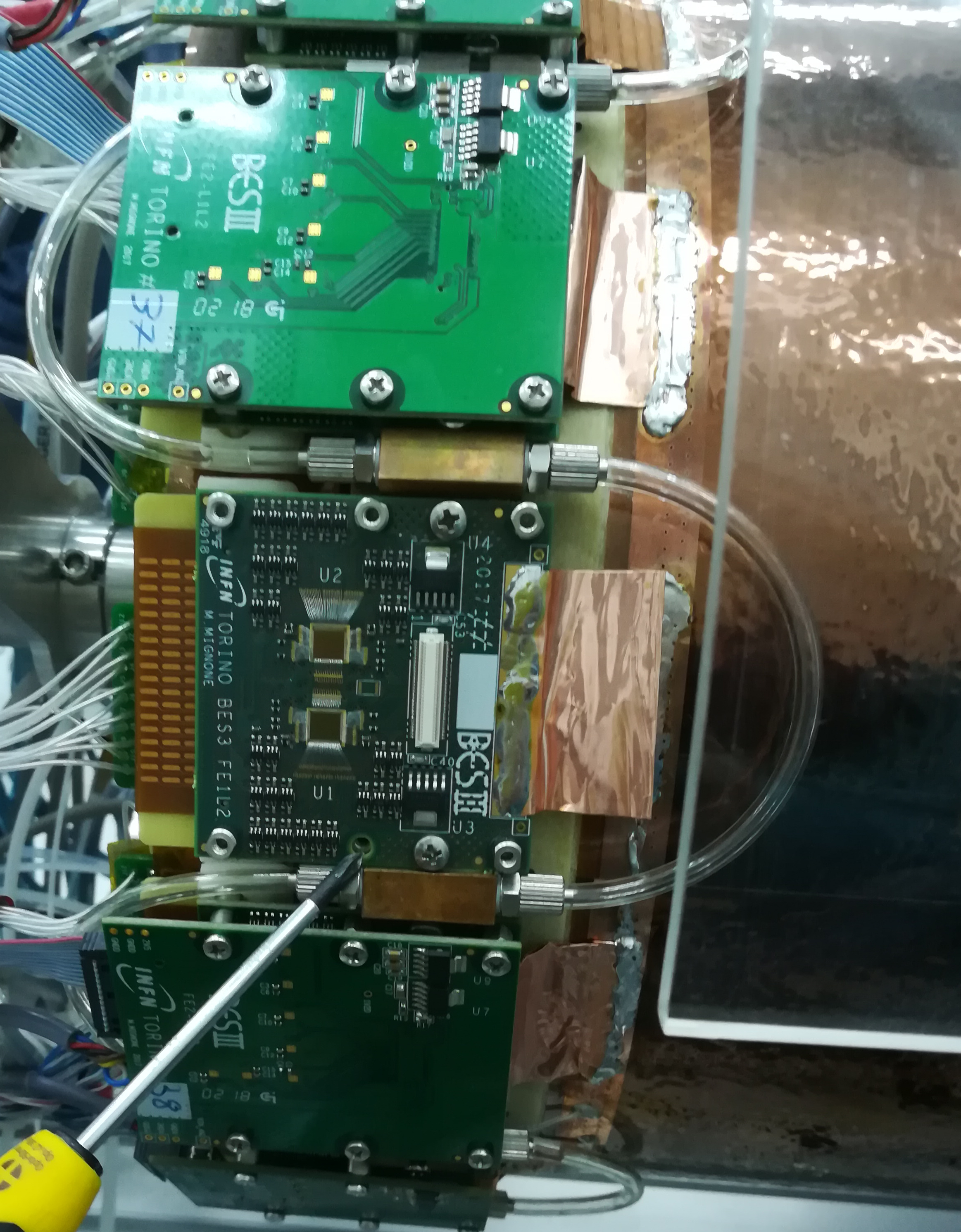}
   \end{subfigure}%
      \caption{\changemarker{On the left: a cut-off of the CGEM detector technical sketch. The readout electronics is located on both sides of the detector. On the right: front-end boards mounted on the detector. One board is removed to show the TIGER ASICs. A copper foil connects the front-end boards ground plane with the detector ground plane.}}
         \label{fig:CGEM_cad}
\end{figure}
\begin{table}
\begin{tabular}{cccccc}
Layer & Inner diameter (mm) & Outer diameter (mm) & Active area length (mm)\\ \hline
1     & 153.8               & 188.4               & 532 \\
2     & 242.8               & 243.4               & 690  \\
3     & 323.8               & 358.5               & 847  
\end{tabular}
\caption{ \changemarker{Dimensions of the CGEM-IT layers. The active area length is equivalent to the length of the $\Upphi$ strips.}}
    \label{table:cgem_sizes}
\end{table}
The CGEM-IT is made of three coaxial layers of triple GEM (figure \ref{fig:CGEM_cad} \changemarker{and table \ref{table:cgem_sizes}). Each cylindrical detector layer is independently assembled, it has an autonomous gas enclosure and can be operated stand-alone. The front-end electronics, the electrodes connections and the gas inlet and outlet are located on both sides of the cylinders}. A GEM electrode is composed of a polyimide foil (\SI{50}{\micro \meter}) coated on both sides with a thin copper layer (\SI{5}{\micro \meter}), pierced with a high density of holes (\SI{50}{\micro \meter} diameter) with a pitch around \SI{140}{\micro \meter}. Applying some hundreds of volts  \changemarker{(\SIrange[]{200}{300}{\volt})} between the copper foils, a 50-100 kV/cm electric field is generated and the electrons accelerated into the holes are thus multiplied \cite{Sauli2016}.\\
In a triple-GEM (figure \ref{fig:CGEM_spac}), the ionisation electrons, produced in the gap between the cathode and the first GEM foil (conversion and drift gap) by the charged particles crossing the detector, are guided by electric fields (transfer fields) through the three GEM foils where they start electron avalanche. Once they cross the last GEM foil they drift to the anode in the so-called induction gap, giving rise to an induced current signal on the anode plane. Several stages of amplification guarantee a high gain of the detector with a low discharge probability.\\ 
The CGEM-IT uses a mixture of argon and isobutane (90\%-10\%).  \changemarker {The GEM voltages and fields settings are shown in table \ref{table:cgem_field}. To set electric fields in the gaps, seven different electrodes need to be biased: the two copper planes of each GEM and the cathode (section \ref{HV}).}
The readout circuit is made by \SI{3}{\micro \meter} copper cladding over \SI{50}{\micro \meter} polyimide. Two arrays of strips are used in order to obtain a 2D readout: \SI{570}{\micro \meter} wide $\Upphi$ strips, parallel to the detector axis, and \SI{130}{\micro \meter} wide V strips, having a stereo angle with respect to the $\Upphi$ strips \changemarker{(see figure \ref{fig:strips})}. The stereo angle is maximized according to each layer active area: \SI{46.7}{\degree} for the innermost layer (layer 1), \SI{-31.0}{\degree} for the central layer (layer 2) and \SI{32.9}{\degree} for the outer one (layer 3).  \changemarker{The pitch, \SI{650}{\micro \meter}, is the same for both arrays and all the layers}.
\\ 
\begin{table} 
\centering
    \begin{tabular}{|c|c|}\hline
    $r\phi$ resolution             &  \SI{130}{\micro \meter} \\ 
    z resolution             &  \SI{<1}{\milli \meter} \\ 
    $\sigma_{p_t}/p_t$         &  \SI{0.5}{\%} at 1 GeV/c\\ 
    Material budget        &  \SI{1.5}{\%} X$_0$ \\ 
    Maximum rate      &  \SI{e7}{\hertz \per \centi \meter^2} \\ \hline
    \end{tabular}
    \caption{Design goals of the CGEM-IT detector.}
    \label{table:cgem_perf}
\end{table}
The full system consists of about 10,000 electronics channels.
Being a tracker, the first goal of the CGEM-IT is a precise position determination, with an expected resolution of \SI{130}{\micro \meter} in \textit{$r\phi$} plane and \SI{300}{\micro \meter} on the \textit{z} coordinate, along the beam direction \cite{marcello2018}. The main design goals are listed in table \ref{table:cgem_perf}. In particular, the CGEM-IT is expected to improve the \textit{z} determination and the secondary vertex position reconstruction with respect to the MDC \cite{GEM_magnetic_field}. The position reconstruction will take advantage from the combination of two different algorithms: 
\begin{itemize}

    \item \changemarker{Charge Centroid extracts the average position weighting the signal amplitude of the firing strips.}
    \item \changemarker{$\upmu$TPC (micro Time-Projection Chamber) uses the drift gap like a time-projection chamber. The positions of the primary ionizations in the drift gap are reconstructed by knowing the drift velocity and the arrival time of the signal at the anode. Then the points are fitted in order to extrapolate the position of the tracked particle in the gap center \cite{Alexopoulos2010,Riccardo2016}.    }
\end{itemize}
These algorithms and the BESIII environment pose specific requirements on the readout features in terms of charge measurement, time resolution and sustainable rate, making the design of a dedicated readout chain necessary.
This paper describes the readout chain of CGEM-IT. The system overview is shown, pointing out the main requirements for the front-end and back-end electronics, including the power distribution systems. 
It is reported how the system has been designed and tested. Finally, a dedicated software interface has been developed in Python with control, monitoring and logging functionalities; its main features are also shown.

\section{System overview}  \label{sec:overview}
\begin{figure}
\centering
  \includegraphics[width=1\linewidth]{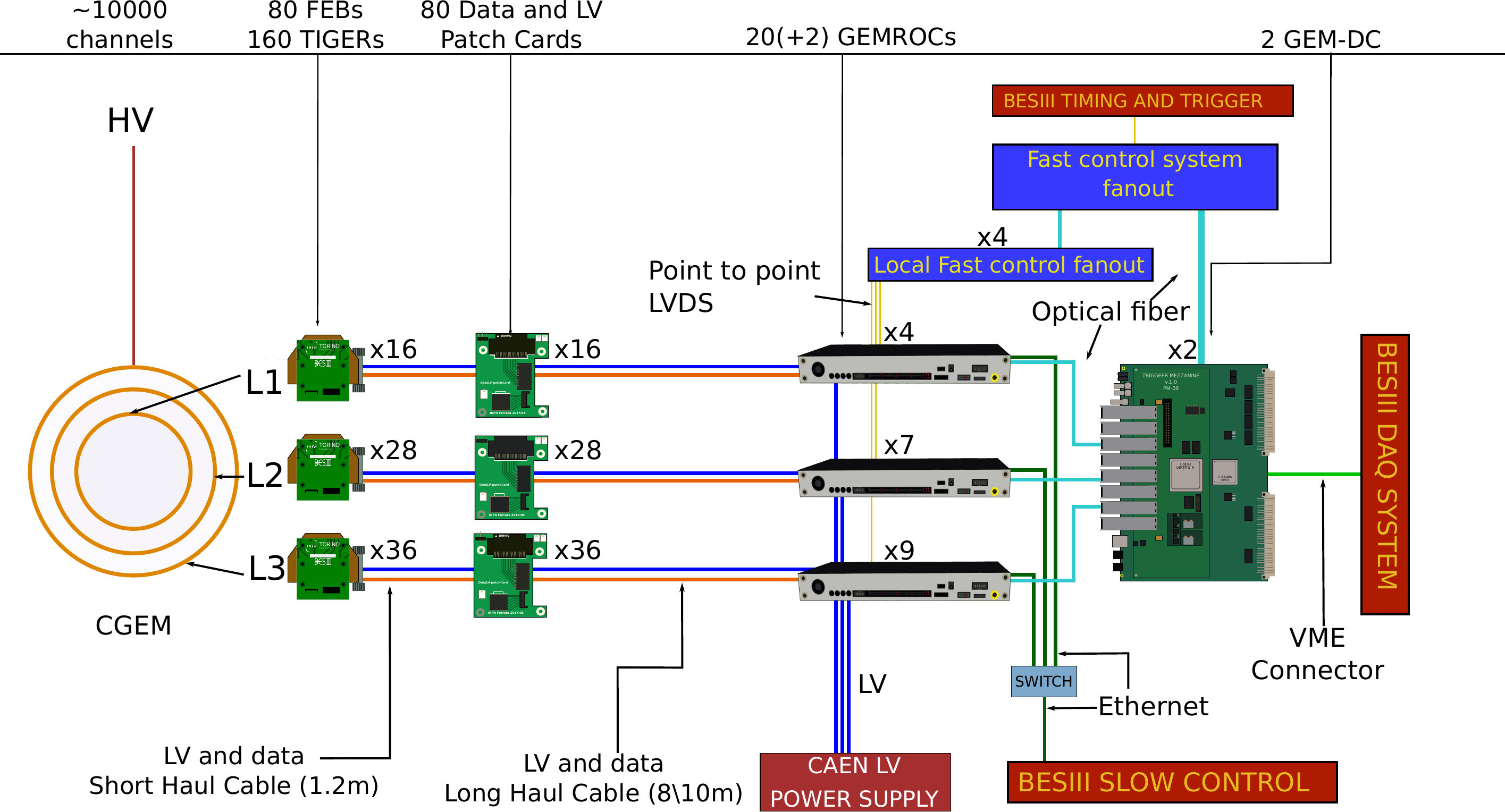}
  \caption{Scheme of the full readout chain.\changemarker{ The $\sim$10000 CGEM-IT detector strips are read by 160 custom ASICs (TIGERs) mounted on 80 front-end boards. The front-end boards are managed, four by four, by the GEM Readout Cards, FPGA based modules which control the ASIC low voltage, configure the ASIC and select the data. The selected data are then collected by two GEM-Data Concentrator cards and sent to the main BESIII DAQ. The fast control system fanout receives the timing signals from the BESIII main system and delivers them to its subsystems.}}
  \label{fig:chain}
\end{figure}

The overall readout chain needs to sustain a peak rate of \SI{14}{\kilo \hertz \per strip} of signal hits for the strips of the innermost layer \cite{design_2014}. In order to ensure that the system will have enough bandwidth and rate capability headroom to accommodate signal and noise, the rate has been multiplied by a factor of safety equal to four,  hence a system with a capability of \SI{60}{\kilo \hertz \per channel} is needed.   \\
Other specific requests were posed by the reconstruction methods: the Charge Centroid requires an analog readout, while the $\upmu$TPC, in order to achieve the needed performance, requires an electronics contribution to the time resolution better than \SI{5}{\nano \second}.\\
The designed readout chain is shown in figure \ref{fig:chain}.\\
The TIGER chips (section \ref{TIGER}) are assembled in pairs on Front-End Boards (FEB) and installed on the detector. Data and ASIC Low Voltage (LV) are fed through Data Low Voltage Patch Cards (DLVPC) by the GEM Read Out Cards (GEMROC). The connection from GEMROC to DLVPC is made through long haul cables (\SI{8}{\meter} and \SI{10}{\meter} long for data and  LV, respectively), while flexible short haul cables (\SI{1.2}{\meter} long) are used inside the spectrometer between DLVPCs and FEBs. \\
The GEMROC boards receive  signals from the BESIII timing and trigger interface, communicate with the BESIII slow control via Ethernet interface and via optical fibers with the GEM-Data Concentrator (GEM-DC) cards,  which build the events and communicate with the VME-based BESIII DAQ. The GEMROC boards also manage the front-end boards power supply and configuration as well as the TIGER output data collection
 (section \ref{backend}).\\
The requirements met for the CGEM-IT project make the electronics also suitable for the readout of other innovative micro-pattern gaseous detectors \changemarker{with similar signal characteristics}. For such reason, the whole readout chain was designed keeping in mind strong adaptability and modularity.

\section{Front-end electronics: TIGER ASIC}\label{TIGER}
\begin{figure}
\centering
  \includegraphics[width=1\linewidth]{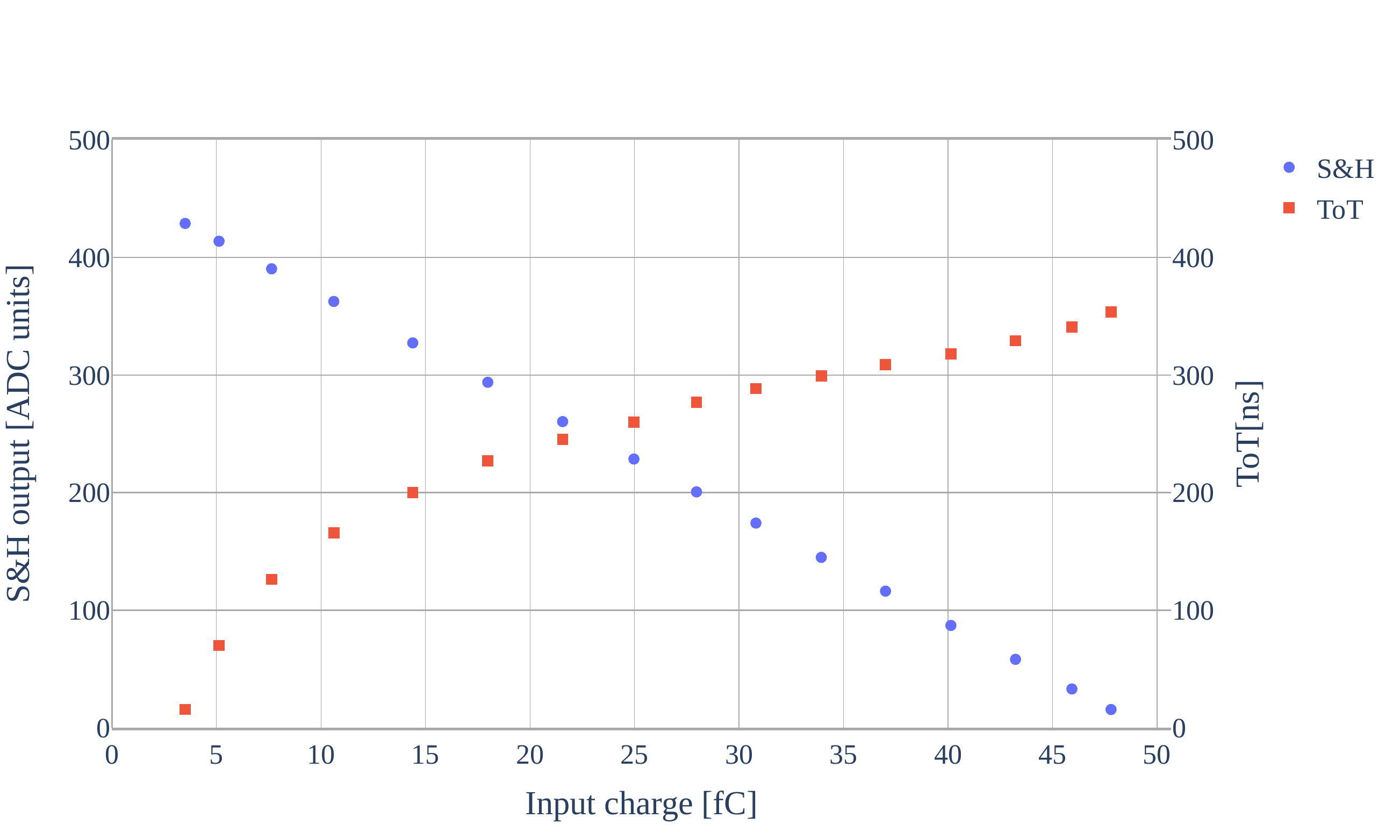}
  \caption{\changemarker{Sample-and-Hold (S\&H) and Time-over-Threshold (ToT) calibration curves for one TIGER channel.} }
  \label{fig:efine_vs_tot}
\end{figure}

TIGER (Torino Integrated GEM Electronics for Readout) is the ASIC designed for the readout of the CGEM-IT strips \cite{RIVETTI2019181}. Each mixed-signal chip can handle the complete readout of the data incoming from 64 channels, providing time and charge measurement in order to satisfy all the design requirements (table  \ref{table:cgem_perf}) .\\
\begin{figure}
\centering
  \includegraphics[width=0.95\linewidth]{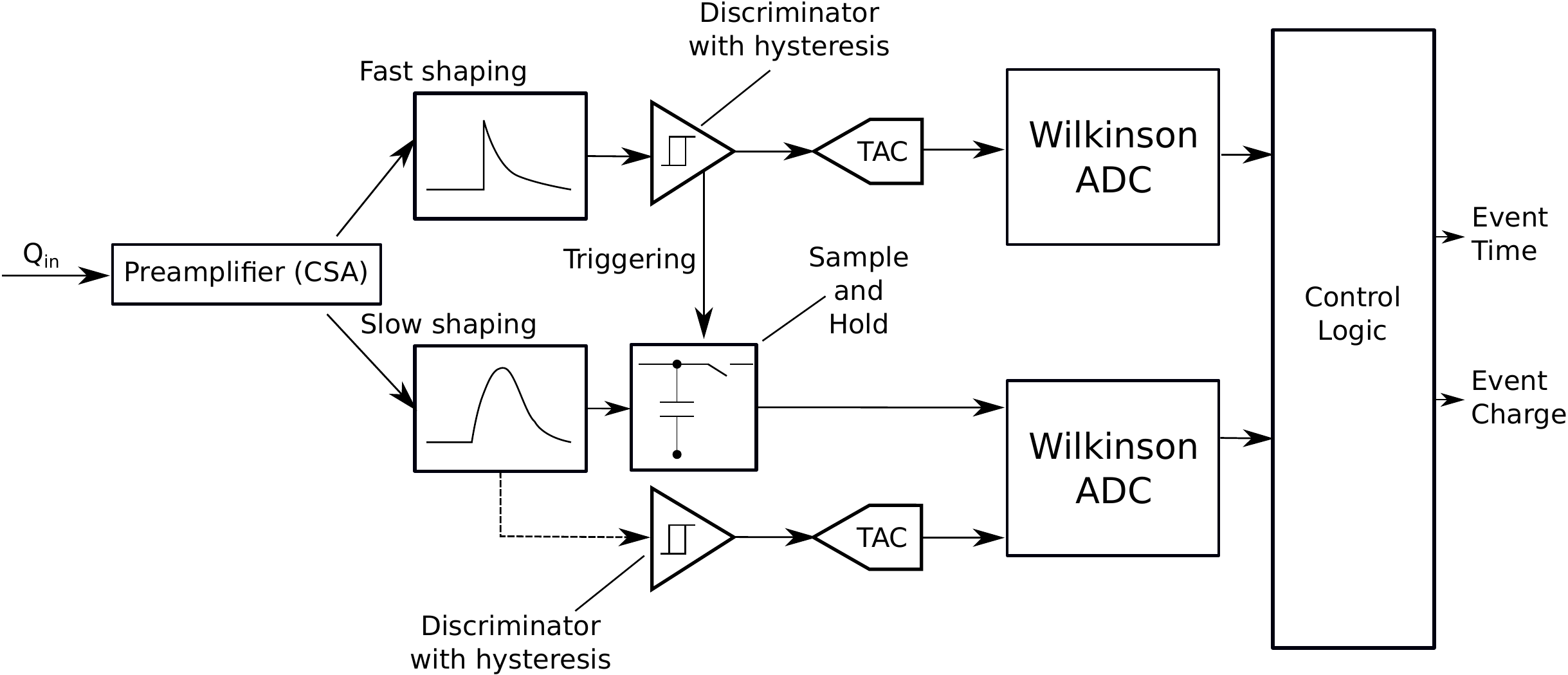}
  \caption{Overview of TIGER \changemarker{channel architecture. After the preamplification, the signal is split in two branches optimized for time and charge measurements. Each branch is equipped with TACs and Wilkinson ADCs for time measurement. On the slow shaping branch (the E branch), the Wilkinson ADC can be used to measure the charge stored by the Sample-and-Hold.}}
  \label{fig:scheme}
\end{figure}
On each channel (figure \ref{fig:scheme}), the signal is amplified and inverted by a three stage cascoded common source Charge Sensitive Amplifier \cite{Geronimo}. The signal is then duplicated to feed two different shaping branches, one optimized for time measurements (\SI{60}{\nano \second} peaking time) and the other optimized for charge measurements (\SI{170}{\nano \second} peaking time). The fast rising time on the time-optimized branch provides low jitter time measures (T branch), while the flatter peak on the charge-optimized branch allows to sample the peak voltage with optimized Equivalent Noise Charge (ENC) for accurate charge measurements (E branch).\\ 
On both branches, a fine time measurement can be performed using a time to amplitude converter combined with a Wilkinson ADC, pushing the time resolution below the clock period (\SI{6}{\nano \second} for BESIII). The jitter on the T branch dominates the final resolution (see table \ref{table:perf}), while the resolution of the TDC is better than \SI{50}{\pico \second}.\\ 
TIGER does not need an external trigger to operate, since it has one discriminator on both branches. The great flexibility of the digital part of the ASIC allows to use many combinations of the discriminator triggers to validate the signals.  \\
The thresholds on the branches discriminators can be set independently for each branch and each channel with 6 bits DACs, while the control logic enables to set various hit validation triggers using both discriminators.\\
The charge can be measured in two different modes: Time-over-Threshold and Sample-and-Hold mode. With the Time-over-Threshold mode, TDCs of both branches on each channel are used to measure the rising and the falling edges of the signal, thus obtaining the signal amplitude. The time information can be acquired on both shapers' output using different configurations.\\
In Sample-and-Hold mode, the signal on the E Branch is stored on a capacitor at a settable time after the threshold crossing, in order to sample its peak value. Then, the stored signal is digitized using the Wilkinson ADC shared with the TDC, providing a linear relationship between the digital output and the input signal amplitude. \changemarker{The comparison between Sample-and-Hold and Time-over-Threshold typical calibration curves is shown in figure \ref{fig:efine_vs_tot}. The Sample-and-Hold mode is the one chosen for the CGEM-IT readout since it is more robust to differences of the detector signals duration. Its calibration is linear and less threshold-dependent than the Time-over-Threshold one.}\\
The ASIC outputs three kinds of data words: hit words, containing the hit information in 54 bits; counter words, used for debugging to count the hits on a channel; and frame words, sent every $2^{15}$ clock cycles, for time reference.\\
The ASIC digital back-end and the full-digital interface are inherited from the TOFPETv2 ASIC \cite{DIFRANCESCO2016194}. Four TX LVDS links can transmit the data in 8b/10b encoding, up to \SI{200}{\mega \hertz} in both Single Data Rate and Double Data Rate, while ASIC internal registers  are programmed via a \SI{10}{\mega \hertz} SPI-like configuration link. In TIGER, triple redundancy of the digital registers has been added to ensure Single Event Upset protection to operate in a high radiation environment.\\
The ASIC performance is summarized in table  \ref{table:perf} \cite{fabio_tesi}.\\
\begin{table}
\centering
\begin{tabular}{|c|c|}\hline
Input dynamic range            &  2-\SI{50}{\femto \coulomb } \\ 
Gain (E branch)             &  \SI{11.8}{\milli \volt \per \femto \coulomb} \\ 
Noise (E branch)            &  < 1800 $e^-$ ENC (\SI{0.29}{\femto \coulomb}) \\ 
Jitter (T branch)           &  < \SI{4}{\nano \second}\\  
Sample-and-Hold residual non linearity &  < 1\% in the whole dynamic range \\ \hline
\end{tabular}
\caption{TIGER key performance measured on silicon with $C_{in}=$\SI{100}{\pico \farad} and $Q_{in}=$\SI{10}{\femto\coulomb}.}
\label{table:perf}
\end{table}
The ASICs are mounted in pairs on front-end boards. Such boards are composed of two different PCBs, mounting a Hirose FX10A144P connector for the interface with the detector strips and the circuitry necessary for the power supply, the biasing, the channel protection and the ESD protection.\\
A cooling system has been designed to stabilize the FEB temperature, to ensure an operation of the ASIC at an almost constant temperature and to limit the heat dissipation towards the outer MDC.  Each FEB mounts a custom designed copper heatsink.  The delivery and the exhaust are made through polyurethane radiation-resistant tubes. The heatsinks are grouped in order to limit the pressure drop per layer. In BESIII, the  CGEM-IT cooling system will be connected to that of the EMC, using a booster pump and a patch panel with three lines in and three lines out per detector side. The cooling system keeps the temperature of the FEBs between \SI{25}{ \celsius} and \SI{30}{ \celsius} by circulating chilled water at \SI{20}{ \celsius}, \SI{2.5}{\litre / \minute} and \SI{2}{\bar}. The thermal load is about \SI{3.2}{\watt} per FEB (half for the ASICs and half for the voltage regulators), for a total of \SI{256}{\watt} for the whole system.

\section{Back-end electronics}\label{backend}
GEM Read Out Cards (GEMROC) are the modules designed to configure and read TIGERs.
Each GEMROC handles four FEBs, for a total of eight TIGERs. The core of each GEMROC is a development kit based on an  FPGA of the Intel/ALTERA ARRIA V GX family  \cite{ARRIVA}, which is connected to an interface card designed for the BESIII experiment.\\
\begin{figure}
  \centering
  \includegraphics[width=0.8\linewidth]{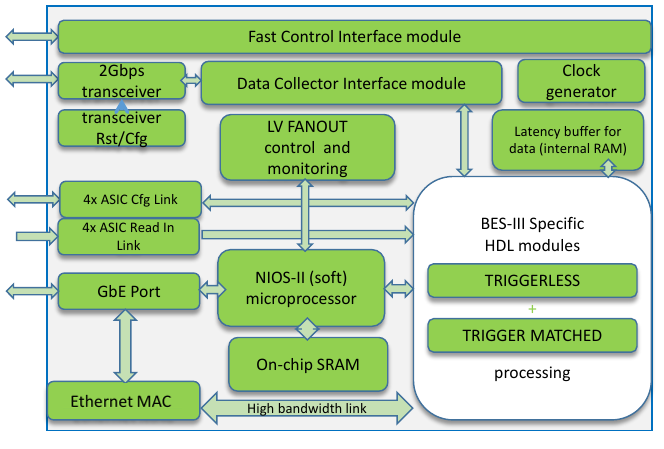}
  \caption{\changemarker{FPGA firmware blocks design. The I/O links are shown together with the main functional blocks. The system integrates both FPGA HDL blocks, in charge of the data processing and I/O managing, and a NIOS-II processor, used to manage the low voltage monitoring and control, the system diagnostic and the GEMROC configuration.}}
\label{fig:gemroc}
\end{figure}
The GEMROC modules distribute digital and analog supply voltages to the FEBs, monitoring their supply currents and operating temperatures, via thermistors installed on the FEB, to ensure safe operation of the on-detector electronics. The whole system can be readout using 20 modules, but, to assure a symmetric distribution scheme of the two sides of each layer, 22 will be used.\\
The GEMROC modules receive, through a suitable distribution system, the timing signals forming the BESIII Fast Control System:
\begin{itemize}
    \item \textit{Clock}: BESIII distributes a \SI{41.65}{\mega \hertz} clock, corresponding to the radio-frequency of the BEPCII storage ring divided by 12. The GEMROC derives all the time references from this signal. The GEMROC FPGA uses internal PLLs to derive signals toggling at a frequency of 166.6 MHz, four times the BESIII clock rate, to drive the TIGER clock inputs.
    \item \textit{L1 trigger}: the BESIII level 1 (L1) trigger is used by BESIII DAQ and by the CGEM-IT readout system to mark event data to be saved. It has a rejection rate around $1:10^4$, a latency, with respect to the event, fixed at \SI{8.6}{\micro \second}, an acceptance window of \SI{1.6}{\micro \second}, an average frequency of \SI{4}{\kilo \hertz}  and a dead time of \SI{3}{\micro \second}. The trigger signal lasts 8 clock cycles \cite{trigger}.
    \item \textit{Check}: every 256 L1 triggers, a \textit{check} signal is sent to verify the subsystems synchronization. This line can also be used, when operating in standalone mode, by the GEMROC modules for debugging purpose (\textit{e.g.} to synchronize the test pulse generation).
    \item \textit{Full}: this signal is used by all BESIII subsystems to notify that the buffers in which the event data are held, pending transmission to the DAQ, are filling up. When the BESIII Fast Control System receives the FULL signal from any of the sub-detectors, it stops sending further L1 triggers.
    \end{itemize}
The FPGAs are programmed to control the system data acquisition; the firmware blocks are shown in figure \ref{fig:gemroc}. \changemarker{The firmware is written partly in Verilog and in VHDL languages}. The GEMROC modules can handle the data received from the FEBs in two different ways. In the first one, called trigger-less (TL), used in standalone configuration for debugging purposes, the data received from the enabled TIGERs are merged and transmitted over the Ethernet output port using the UDP protocol. A UDP packet, limited to the standard size of \SI{1500}{B}, is transmitted either when it contains all data collected in eight TIGER time frames ($2^{15}$ TIGER clock periods) or when the number of TIGER collected data words reaches 180, with the remaining part of the TIGER data being transmitted in following UDP data packets. \\
Differently, in trigger-matched (TM) mode, a finite state machine selects the hits to be sent over in the following way: the data incoming from each couple of TIGERs are stored in a \textit{latency buffer} circular memory, which is organized in pages of 32 locations each. This memory is intended to buffer the incoming TIGER data, pending the L1 selection trigger.  A latency buffer page, also defined bucket, contains all the data recorded by the two TIGERs of a FEB in a time interval corresponding to $2^8$ TIGER clock cycles (\SI{1.53}{\micro \second}). The bucket memory is circular, so data are overwritten every address rollover (\SI{24.6}{\micro \second}). The paged organization of the latency buffer is meant to speed up the search, started when the L1 trigger signal is received.\\
When the BESIII trigger arrives, the FPGA logs the trigger time of arrival timestamp, waits for a programmable delay to
account for the stochastic transmission latency of the data over the TIGER output serial links, and then reads the buckets determined by the L1 trigger time of arrival timestamp to search the TIGER data with coarse event timestamps falling  inside the BESIII trigger window. The hits from all the FEBs connected to a GEMROC and enabled to data taking are merged to prepare a trigger-matched data packet to which a header and a trailer are added. The L1 trigger to which the data were matched can be properly identified through the header and trailer. The latter contains also diagnostic information about the GEMROC status.\\
The trigger-matched data packet is then sent over the fiber optic link to the GEM-DC modules and also sent as a UDP packet over the Ethernet port.\\
This last feature is exploited during standalone operation of the CGEM-IT setup and enables trigger-matched data to be also collected by an alternative Ethernet-based data acquisition system.\\
The GEM-DC are VME 6U cards used to collect the trigger-matched data packets transmitted, via optical links, by the GEMROC modules and to assemble them into sub-detector events identified by the common trigger number.
A VME interrupt is then generated by the GEM-DC to prompt the VME crate CPU to read the trigger-matched events stored in the GEM-DC buffers. The GEM-DC boards inherit the hardware design and most of the firmware from the Read-Out Driver (ROD) modules used by the KLOE-2 inner tracker \cite{DAQ_KLOE}. \\
Each FEB is connected to a GEMROC via short-haul and long-haul shielded multiple twisted-pair (MTP) cables interconnected through the DLVPC boards. The MTP cables feature relatively low losses but, due to the length of the data path and the unavoidable discontinuities, the fast signals exchanged between GEMROC and FEB, and the TIGER clock in particular, have been found susceptible to EMI and ground noise. This causes transmission errors that are detected thanks to the error detecting features of the 8b/10b encoding of the TIGER serial outputs.\\
To improve the signal integrity issues, a set of patch cards have been installed on the interface card of the GEMROC modules; these cards feature an LVDS driver with pre-emphasis, which boosts of about \SI{20}{\percent} the amplitude of the TIGER clock signal received by the FEBs, resulting in a signal level well matching the input specifications of the clock receivers on the FEBs. \\
To improve the quality of the fast control signals (FCS) and overcome the signal integrity issues, an upgrade of the fast control signal distribution system has been carried out. The upgraded FCS distribution system consists of a System FCS Fanout module driving four Local FCS Fanout modules, located near the BESIII DAQ system and near the CGEM-IT GEMROC modules respectively. The GEMROC modules are grouped at four locations around the BESIII detector. Therefore, four Local FCS Fanout modules are being assembled.\\
The System FCS Fanout is interfaced directly to the BESIII FCS System and it is connected to the Local FCS Fanout by bidirectional optical links operating at frequency from DC to \SI{50}{\mega \hertz}. The Local FCS Fanout modules use transceiver daughter cards to drive the received electrical signals onto a multi-drop backplane with single-ended signaling. Up to six Flat-Cable Port (FC-Port) daughter cards are also installed on the Local FCS Fanout backplane. The FC-Port cards feature a 10-pin connector for the flat cable segments onto which the output LVDS signals are delivered to the destination GEMROC module.\\ 
The System FCS Fanout is programmable and may simulate the BESIII timing signals to test the CGEM-IT in standalone mode.
\section{Low Voltage distribution}
The power distribution design is based on the following FEB bias requirements: \SI{1.08}{\ampere} at \SI{1.4}{\volt} for the ASIC analog part, \SI{0.32}{\ampere} at \SI{2.5}{\volt} for the ASIC digital circuitry. A further requirement for the LV power system is to provide the GEMROC modules with a bias voltage of \SI{15}{\volt} with an average current consumption of about \SI{1.1}{\ampere}.\\
The power supply system uses the following commercial modules provided by CAEN S.p.A.:
\begin{itemize}
    \item One SY4527LC mainframe \cite{SY4527LC} (\SI{600}{\watt} max) and three A2517 boards \cite{A2517} (24 channels, max \SI{5}{\ampere/ channel} and \SI{50}{\watt/channel}) to supply the FEBs;
    \item One SY5527 BASIC  mainframe \cite{SY5527} (\SI{600}{\watt} max) and four A2519 boards \cite{A2519} (32 channels, max \SI{5}{\ampere/ channel} and \SI{50}{\watt/channel}) to supply the GEMROC modules.
\end{itemize}
One pair (analog and digital supplies) of output channels of the CAEN A2517 is parallel-connected to the FEB power input of two GEMROC boards, which, in turn, deliver and monitor the analog and digital power supplies to eight FEBs.\\
Such modularity of eight was chosen because it well matches to the dimensions and complexity of the GEMROC modules and allows to keep a separate ground reference for each detector layer when 22 GEMROC modules are used.\\
 All power supply cables are shielded and equipped with connectors to  refer shields to the ground net. The length and the current capacity of the cables used for the LV power distribution are hereafter detailed:
\begin{itemize}
    \item from the mainframe rack to the GEMROC:
    \begin{itemize}
        \item GEMROC supply cable: \SI{17}{\meter}, carrying (at most) \SI{2}{\ampere};
        \item FEB supply: \SI{17}{\meter}, carrying (at most) \SI{9}{\ampere} (analog+digital);
    \end{itemize} 
    \item from the GEMROCs to the DLVPCs:  \SI {10}{\meter}, carrying (at most) \SI{1.1}{\ampere};
    \item from the DLVPCs to the FEBs: \SI{1.2}{\meter}, carrying (at most) \SI{1.1}{\ampere}.
\end{itemize}
Throughout the whole LV distribution system, the IR drop is a concern but the toughest constraints are for the cables which are connected between the CAEN A2517 cards and the GEMROC modules; for these cables the four wires section is \SI{4}{\milli \meter^2}.\\
The power dissipation, considering the optimization of the TIGERs operating conditions, is expected to be \SI{330}{\watt} for the total number of FEBs and \SI{484}{\watt} for the total of 22 GEMROC modules.
It  will be increased by the amount dissipated by the FCS Fanout system, which is estimated in \SI{33}{\watt} for the FCS System  Fanout and \SI{22}{\watt} for the FCS Local Fanout.\\
To leave headroom to the power supply system, an A4532 Power Booster Unit will be installed in the SY5527 CAEN mainframe to add \SI{600}{\watt} to its output power.\\
The LV power distribution system is equipped with hardware and firmware features to allow remote control at the level of the CAEN power supply mainframes and at the level of the GEMROC.\\
The CAEN boards control and monitor their individual output channels which provide analog and digital power to the groups of eight FEBs, while the GEMROCs distribute, control and monitor the power to the individual FEB. The LV power distribution system is managed, in standalone operation, by custom software at the FEB level and by the CAEN GECO system (see section \ref{sec:software}) at the power supply mainframe level. Once installed the CGEM-IT inside the BESIII spectrometer, the whole system will be managed by the Detector Control System (DCS) \cite{Ablikim2010}.
\section{High Voltage distribution}\label{HV}
\begin{figure}
\centering
  \includegraphics[width=0.5\linewidth]{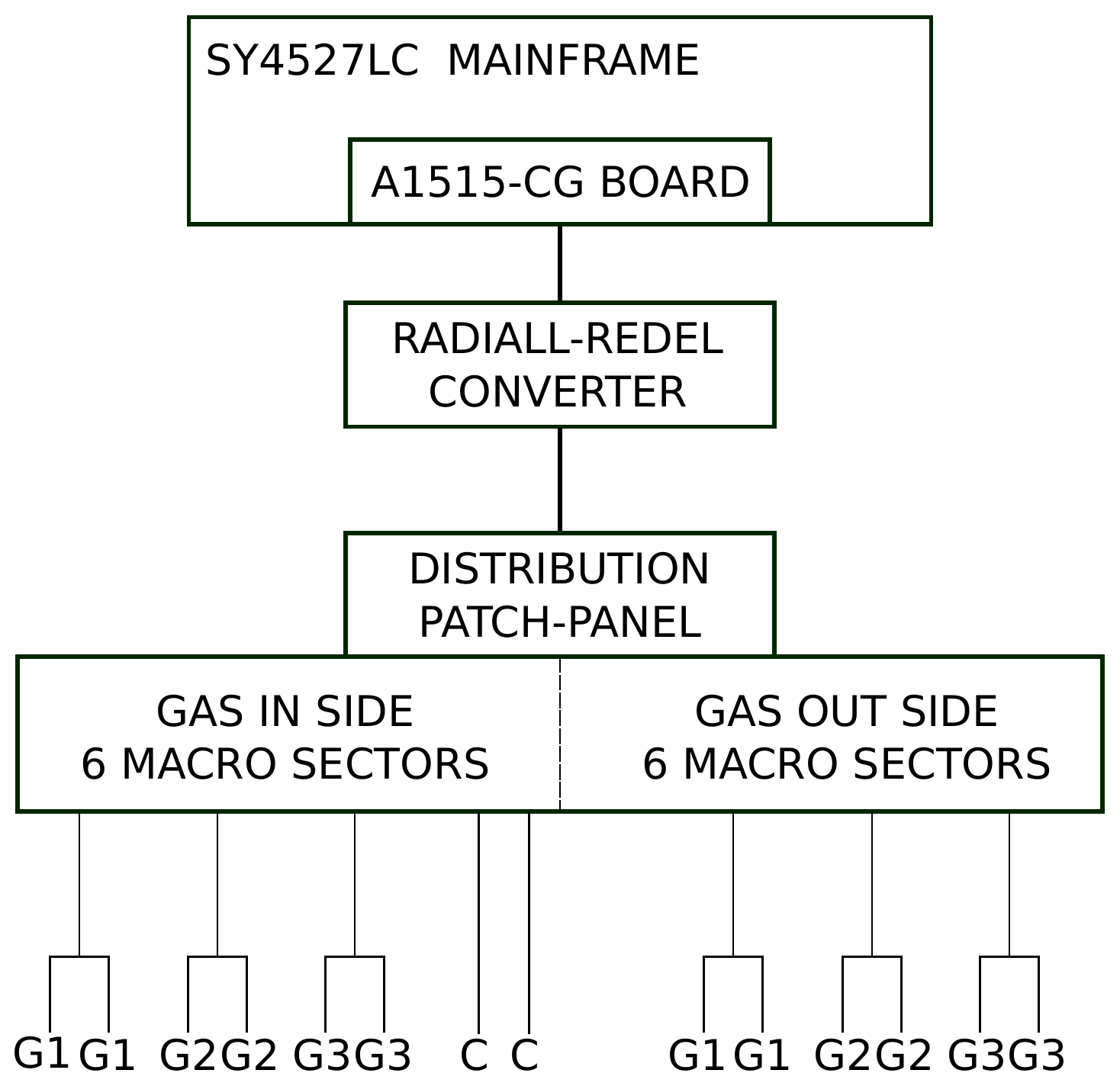}
  \caption{\changemarker{Block diagram of the layer 1 HV distribution system. Each multi-core cable departing from the distribution patch-panel comprises two macro-sectors cables, the 20 corresponding micro-sector cables and six reference ground cables. The power line splits in the two macro-sector cables for the last meter before the detector.} }
  \label{fig:HV_scheme}
\end{figure}

The description of the High Voltage chain is included to complete the picture of the power distribution systems and their interconnection.
For each triple-GEM, seven different electrodes need to be biased (the two copper planes of each GEM and the cathode). One side of each GEM foil is segmented in macro-sectors, the other in micro-sectors. \changemarker{Each macro-sector is matched by 10 micro-sectors. The cathode electrode is not segmented.} The subdivision of the electrodes allows to reduce the energy involved in the case of a GEM discharge and to disconnect a part of the detector in the event of a short-circuit between GEM foils.\changemarker{The number of sectors depends on the layer size: for each GEM, there are 4 macro-sectors for layer 1, 8 for layer 2 and 12 for layer 3. The layer 1 electrodes are etched on a single sheet, while those of layer 2 and layer 3 on two sheets, glued together.\\
The HV distribution for layer 1 is shown in figure \ref{fig:HV_scheme}. The electrodes are powered from both sides of the cylinder ("gas in" and "gas out"), in order to optimize the space.}\\
CAEN A1515CG boards \cite{A1515} will be used as main generators, hosted in a SY4527LC  mainframe \cite{SY4527LC}. CAEN A1515CG are boards designed to supply triple GEM detector, providing seven floating outputs on the corresponding detector electrodes. The HV reference ground is connected through a \SI{10}{\kilo \ohm} resistor to the ground plane of each layer.
The boards are interfaced through a Radiall-REDEL converter to a custom passive distribution patch-panel that splits the seven inputs into all the levels necessary to power the detector. Each output connector supplies two macro-sectors and the related micro-sectors. A full panel supplies up to 18 macro-sectors and the related micro-sectors. The panels allow to disconnect single sectors for diagnostic purpose or to isolate a shorted micro-sector.\\
The supplies, the Radiall-REDEL converter and the distribution panel will be hosted on the platform located on top of the spectrometer. 
For the off-dectector routing, braid shield and halogen free cables, rated up to \SI{4.5}{\kilo \volt}, will bring the power from the patch panels to the HV connectors cards inside the spectrometer. The appropriate shielding of these \SI{18}{\meter} long cables is essential to lower the HV induced pick-up noise of the system. The \SI{2}{\meter} long cable for on-detector interconnections has to be light, multi-core and flexible to allow routing in tight spaces. Custom small-section and low-weight cables have been assembled using extruded FEP insulated cables. For the same reasons of size and weight, custom connectors for off-detector/on-detector connections were designed. The connectors are composed of two boards which are inserted one into the other. The electrical insulation is ensured by a protective coating and by 3D printed plastic enclosures.\\
The connectors towards the CGEM-IT are similar to those of interconnections. Since the tight space does not allow to install any enclosure, the electrical insulation is ensured by the board clearance and by the insulating coating. They mount protection resistors (\SI{1}{\mega \ohm} for macro-sectors and \SI{10}{\mega \ohm} for micro-sectors). The coupling between the protection resistors and the macro/micro-sectors capacitance provides RC filtering on the HV lines. \\
All power supplies are fully floating and each supply has its own return path. The whole detector is located in a Faraday cage. The detector module is connected to earth at only one point (star structure).
\section{Occupancy and \changemarker{d}ata rate}
As seen in section \ref{sec:overview}, the maximum hit rate per channel for the CGEM-IT detector in the BESIII experiment is about \SI{60}{\kilo \hertz}, considering a factor of safety.\\
The system is well-sized to address the experiment rate. Great flexibility in the delay tuning and the thresholds settings allows to allocate the resources to maximize the performance taking into account the noise and background differences among the layers and the strips.\\
\begin{figure}
\centering
  \includegraphics[width=\linewidth]{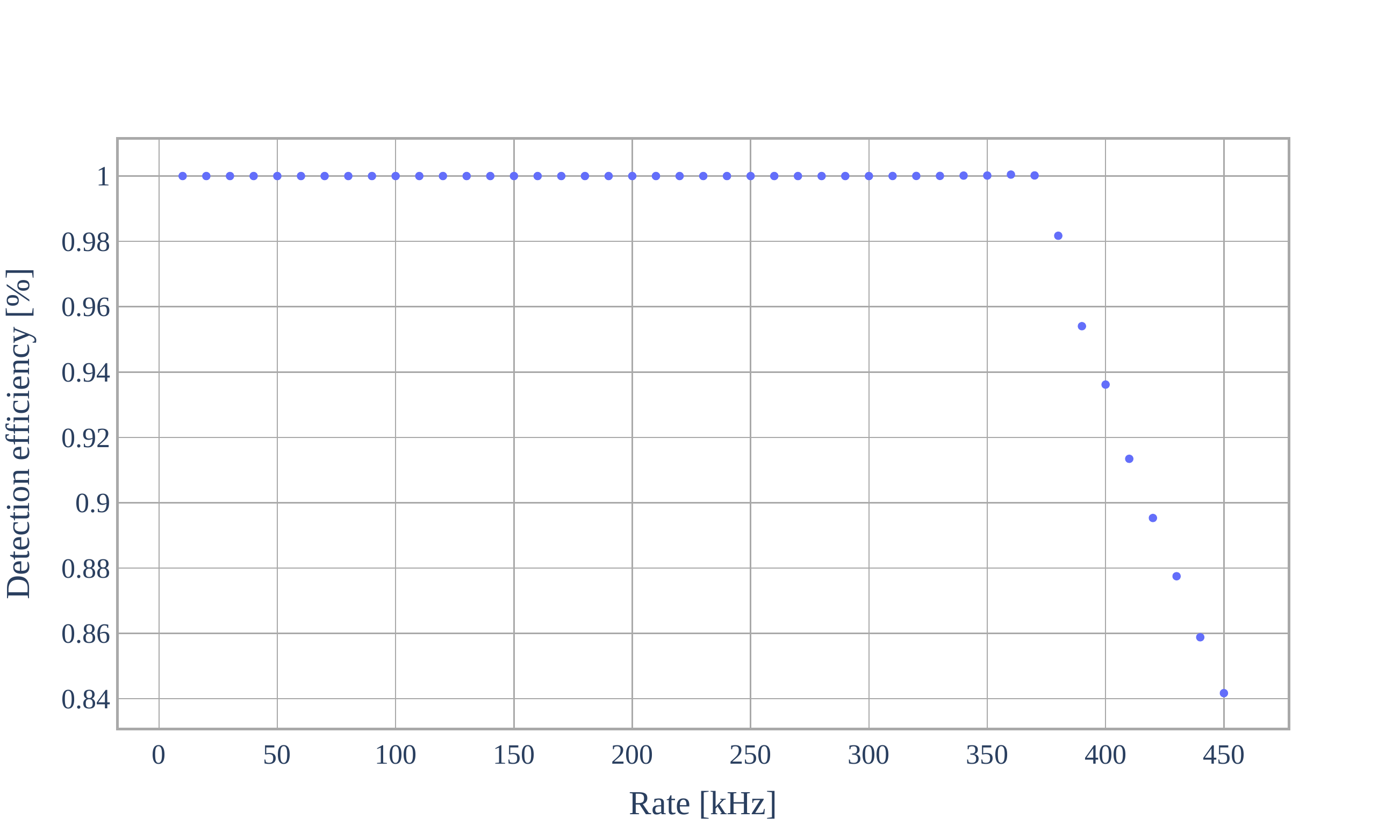}
  \caption{Digitization capability of a single TIGER channel. The measurement is done with an externally generated voltage \changemarker{square waves} injected in the channel under test for 10.000 times at different rates and then the number of digitized hits is counted. The statistical errors are negligible. 
  }
  \label{fig:tige_max_rate}
\end{figure}
\begin{figure}
\centering
  \includegraphics[width=1\linewidth]{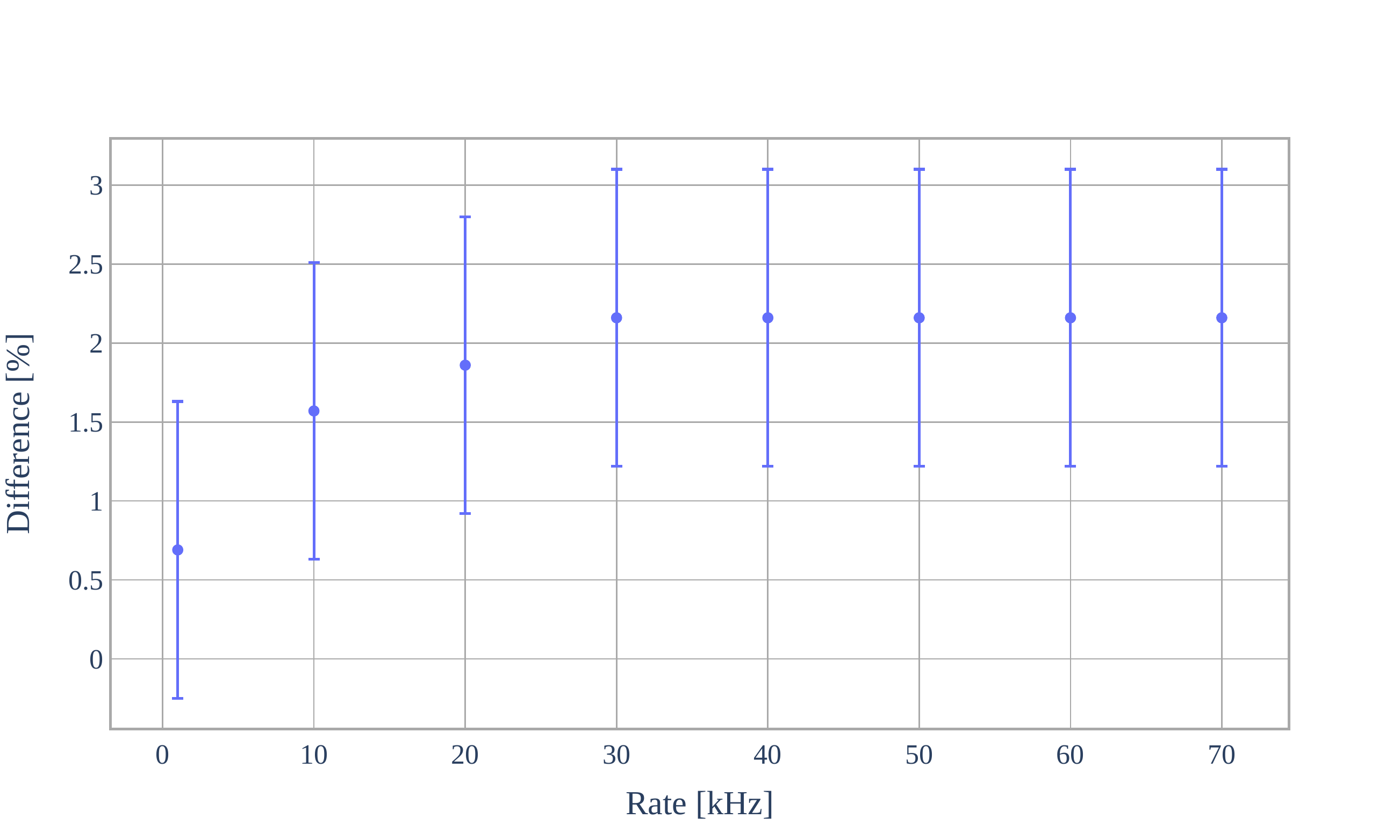}
  \caption{Difference between the charge measured and the \changemarker{charge according to the} calibration, with respect to the rate. It was estimated for a fixed input charge of \SI{40}{\femto \coulomb}, injecting a current pulse in one channel \changemarker{using a voltage generator and a capacitor. To generate an approximate single polarity current signal, the capacitance was rapidly charged (\SI{50}{\nano \second}), and slowly discharged (\SI{14}{\micro \second})}. 
  The error is the standard deviation of the measured charge.}
  \label{fig:tige_max_rate_charge}
\end{figure}
The TIGER ASIC was tested to assess its maximum sustainable hit rate: figure \ref{fig:tige_max_rate} shows that the TIGER can digitize at a hit rate per channel up to \SI{300}{\kilo \hertz}. Figure \ref{fig:tige_max_rate_charge} displays that the ASIC can maintain its accuracy in charge measurements at hit rates per channels up to \SI{70}{\kilo \hertz}, even with large signals of \SI{40}{\femto \coulomb}. \changemarker{For the CGEM-IT operation, where we can expect an average input charge of \SIrange[]{8}{10}{\femto \coulomb} and a physics hit rate of \SI{14}{\kilo \hertz \per strip}, the difference between the charge measured and the charge according to the calibration is within the relative uncertainty (\SI{3}{\percent}) of the Sample-and-Hold noise \cite{fabio_tesi} and should not affect the charge-centroid resolution.}\\
The two LVDS, 8b/10b encoded, serial data links driven by each TIGER at a rate of \SI{332}{\mega b \per \second} enable the transfer of hit data at a maximum lossless rate of \SI{4.15}{\mega \hertz}, equivalent to  \SI{65}{\kilo \hertz} per channel assuming a flat distribution of hits in time. The TIGER performance is thus more than adequate to cope with the BESIII expected data rate.\\
The GEMROC modules have enough resources to accommodate the input and the output data rates.
A set of rate-leveling input FIFOs in the GEMROC FPGA receives and offers temporary storage for the data from each pair of TIGERs on the same FEB. Data are then accommodated into a set of circular page-organized memory buffers, one per FEB, on which they are stored for the L1 latency period (plus some programmable delay to account for transmission latency) until the decision to save or reject them (see section \ref{backend}).\\
At the expected background rate, an average of 12 memory positions will be occupied in each 32-locations page of the circular buffer, leaving a headroom of 20 locations for physics signals and noise, which is sufficient in response to a typical BESIII interaction with an average of six tracks, an average cluster size of 3, on both the $\Upphi$ and the V views on each layer.\\
Unused resources in the GEMROC FPGA are still available to increase the capacity of all storage devices described above if the need arises in the future.\\
The GEMROC output data rate is determined by the BESIII event rate, the L1 trigger rate, the L1 trigger window and the average trigger-matched event size. \\
According to the conservative estimate for the overall physics and background rate of \SI{60}{\kilo \hertz} per TIGER channel, an average hit rate of \SI{3.9}{\mega \hertz} per TIGER is expected. Considering that the L1 trigger acceptance window width is \SI{1.6}{\micro \second} and that each hit information is encoded in an \SI{8}{byte} word, the size of the TM data packet (including header and trailer) sent in reply to a L1 trigger would be around \SI{410}{byte}. Taking into account that the average BESIII L1 trigger rate is \SI{4}{\kilo \hertz}, the data bandwidth for trigger-matched data amounts to slightly more than \SI{1.6}{\mega B \per \second} per GEMROC and thus about  \SI{33}{\mega B \per \second} for the full detector.\\
These bandwidths are achievable by the GEMROC data processing modules and by the optical fiber and Ethernet communication ports.\\
The optical communication links between the GEMROCs and the GEM Data Concentrators operate at \SI{2}{\giga b p \second}. Since the serial data is 8b/10b encoded the resulting net bandwidth is \SI{1.6}{\giga b p \second}, or \SI{200}{\mega B \per \second}, which is well above the TM data rate of \SI{1.6}{\mega B \per \second} discussed above.\\
In the same scenario, the transmission of the average TM data packet over the optical link requires about \SI{2.0}{\micro \second}. The link transmission latency is negligible considering that the GEM-DC starts processing the input data as the TM packet headers are received, overlapping, therefore, data reception and data processing.
Two GEM-DC modules are planned to read the 22 GEMROC, since each GEM-DC features 16 input optical transceiver ports. \\
The GEM-DC supports a wide range of VME protocols over the VME backplane, including 64-bit double-edge asynchronous (2eVME) and synchronous (2eSST) transfers, as well as serial transfer protocols for the 8-lane \SI{3.125}{\giga b p \second} VXS port. Standard block transfers at \SI{40}{\mega B \per \second}, initiated by interrupts from the GEM-DC, are currently used to readout the GEM-DC via the standard VME-based BESIII DAQ.
\section{Control software}
\label{sec:software}
In order to characterize, debug and test the system before the installation, a control software was developed. This software (called Graphical User Frontend Interface - GUFI), written in Python, handles electronics operations and acquisitions via the UDP protocol. The software has an intuitive graphical interface in order to be easy to use and to better control the system multiplicity.\\ 
\begin{figure}
\centering
  \includegraphics[width=0.8\linewidth]{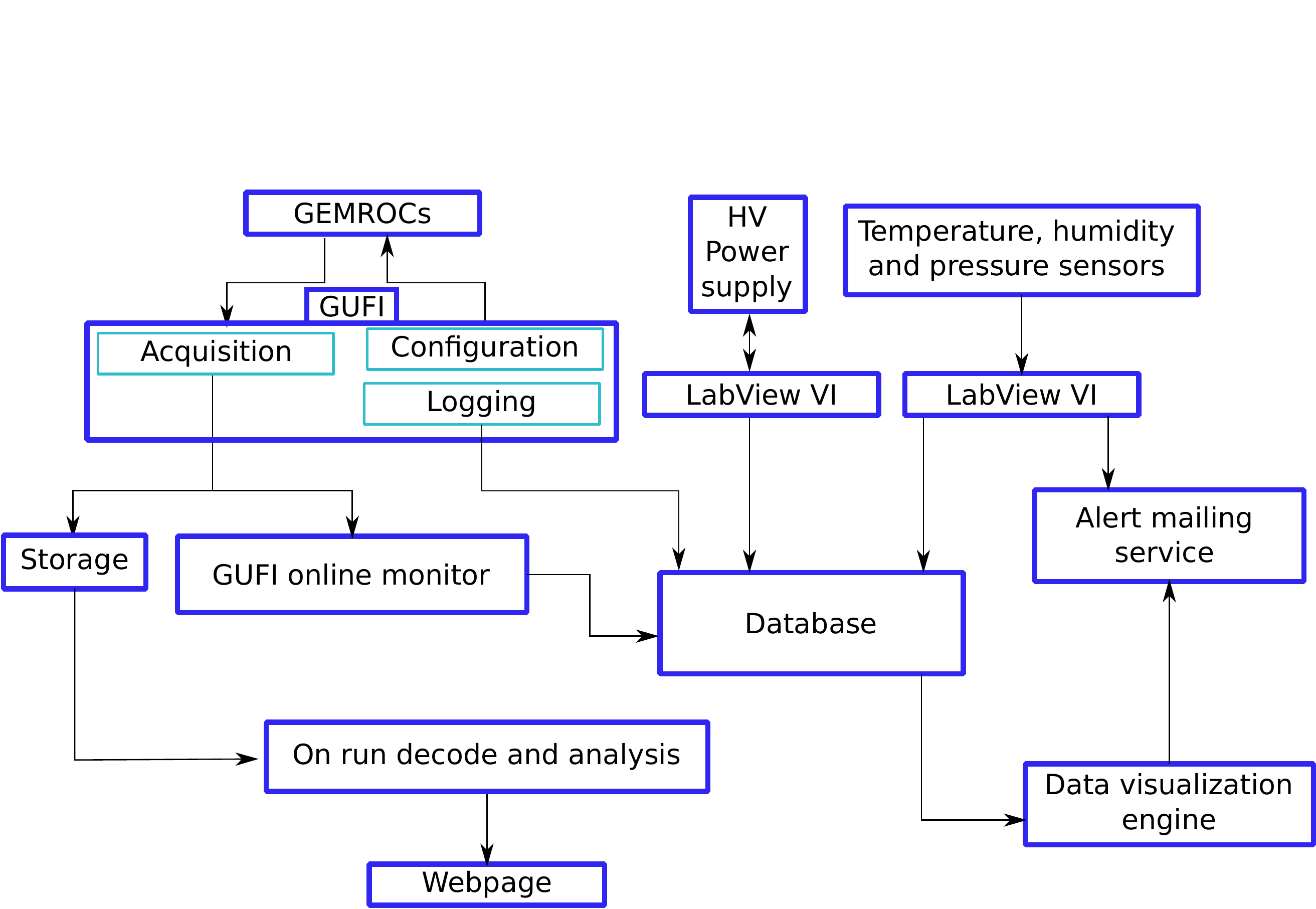}
  \caption{The acquisition and test control software infrastructure. Many parts of the software can be installed and run in different computers, since the communication takes place via UDP or TCP protocol.}
  \label{fig:gufi_scheme}
\end{figure}
Beside the standard operations, (e.g. powering on the front-end electronics, writing the configuration  and starting the acquisition), the software has advanced features:
\begin{itemize}
    \item Scan of the communication delays for the LVDS links between GEMROCs and TIGERs, in order to establish the most robust possible communication;
    \item Threshold-scans in order to verify the channel condition and the noise levels;
    \item Threshold optimization algorithm: the software is capable to scan the 2D thresholds space to obtain a certain rate of noise. The desired rate can be chosen for each part of the system ($\Upphi$ or V strips for each layer), to optimize the homogeneity of the noise;
    \item Diagnostics and slow control operations.
\end{itemize}
The software has also monitoring and logging functionalities. The pieces of information are written in an Influx-DB database \cite{Influx}, which can be queried directly or visualized via the Grafana interface \cite{Grafana}. 
The GUFI online monitoring part creates a copy of each incoming packet and sends them to an analysis software. In this way, it is possible to prioritize the data storage and use the remaining resources for the online data analysis.
Indeed, the software interconnects with the GRAAL analysis software suite \cite{Farinelli_2020}, in order to perform a fast online analysis of key features to assess the status of the detector.\\
The GUFI system, together with the acquisition logging, the environment logging  and the HV status read by LabView VIs, allowed to operate remotely the detectors assembled in China during the COVID-19 outbreak. The scheme of the full software system is shown in figure \ref{fig:gufi_scheme}.\\
The software is very flexible and can be employed in future tests with developments of the detector, other detectors and other readout chains.\\
In order to test the CGEM-IT operation, a system made of two, out of three, fully equipped layers has been set up in a facility at the Institute of High Energy Physics in Beijing. The setup composed of layer 1 and layer 2 is fully equipped with gas, power and cooling systems, readout electronics, and the computing needed to operate it. A cosmic-ray trigger exploiting  plastic scintillator detectors in coincidence was also set up to allow the system to record and fully reconstruct cosmic ray events.\\
This setup has been taking data since December 2019, testing the readout chain reliability and stability, while providing data for the development and the calibration of the analysis software.\\
\begin{figure}
\centering
  \includegraphics[width=0.9\linewidth]{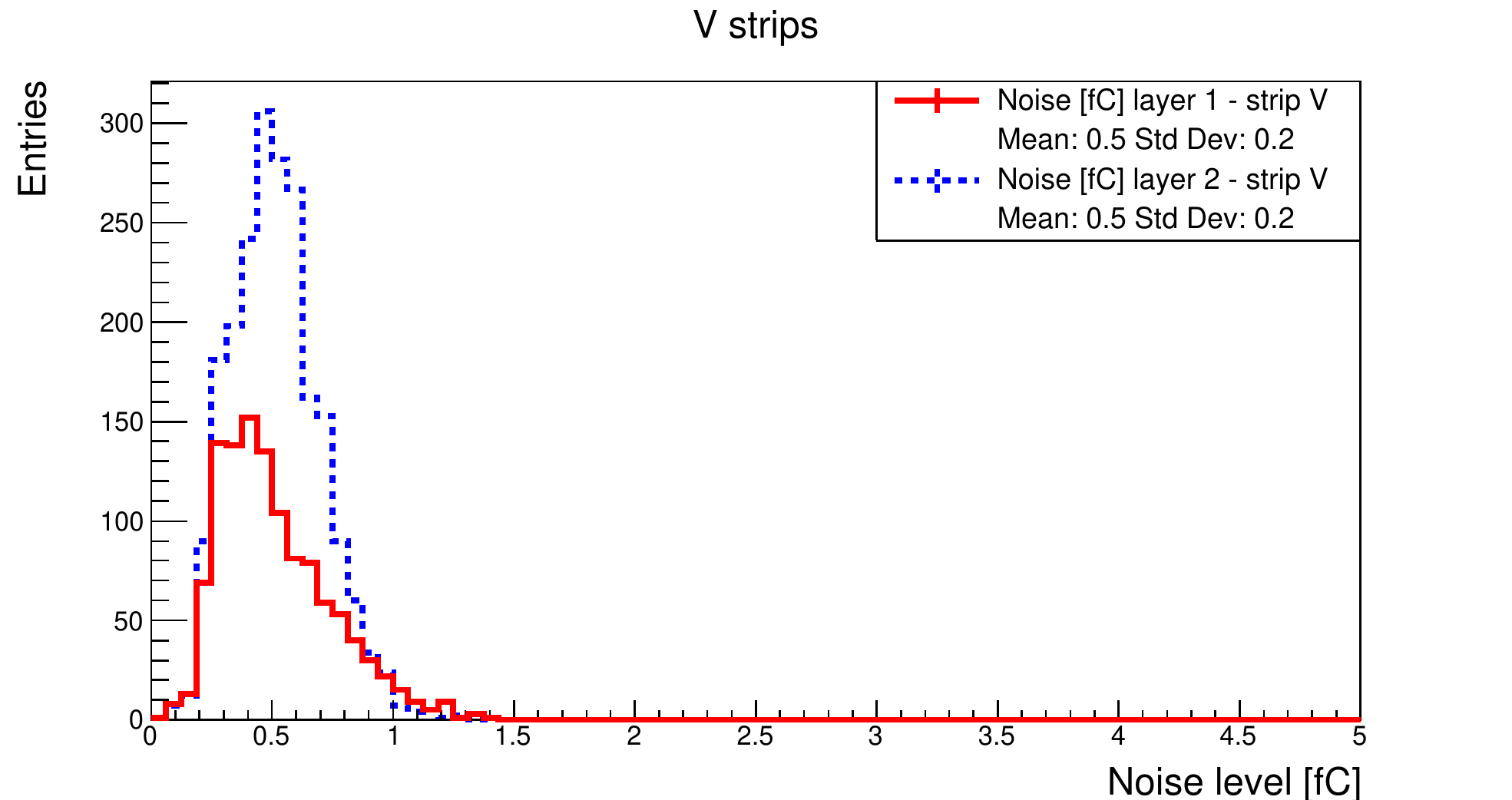}
\centering
  \includegraphics[width=0.9\linewidth]{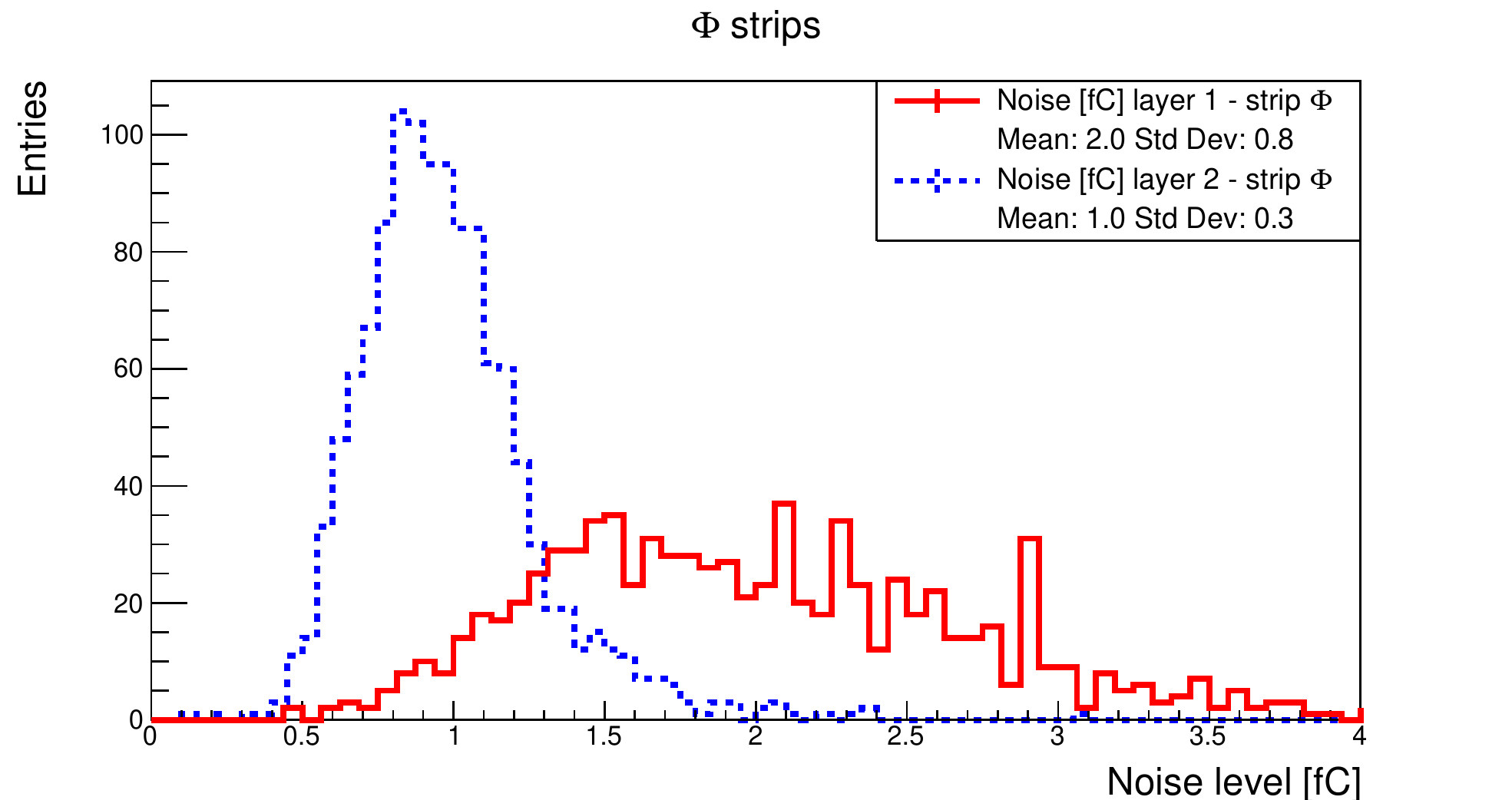}
  \caption{Noise level measured on layer 1 and layer 2, both for \changemarker{$\Upphi$} and V strips, using the cosmic ray test setup in Beijing.}
    \label{fig:noise_x}
\end{figure}
\begin{figure}
\centering
  \includegraphics[width=1\linewidth]{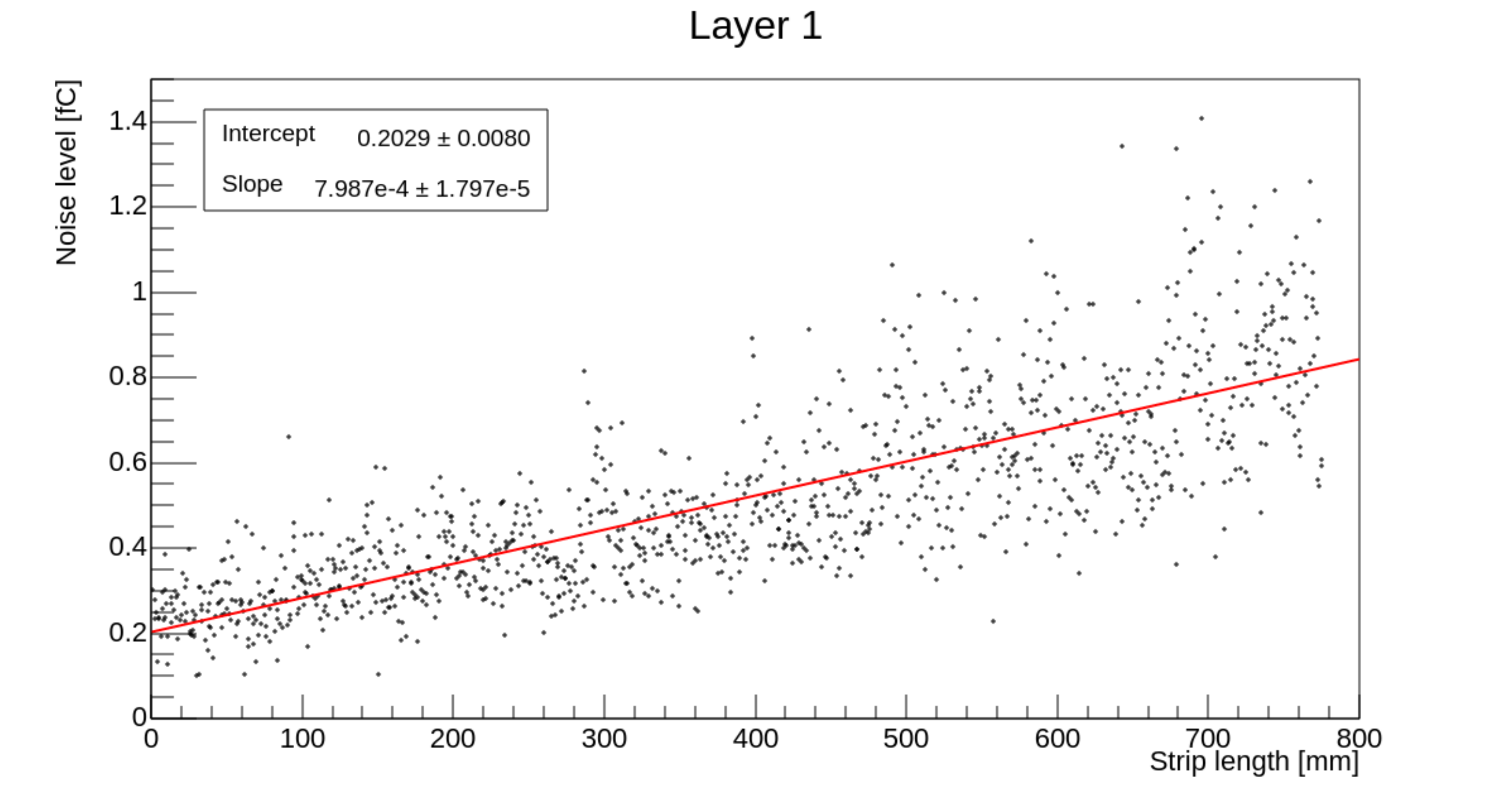}
\centering
  \includegraphics[width=1\linewidth]{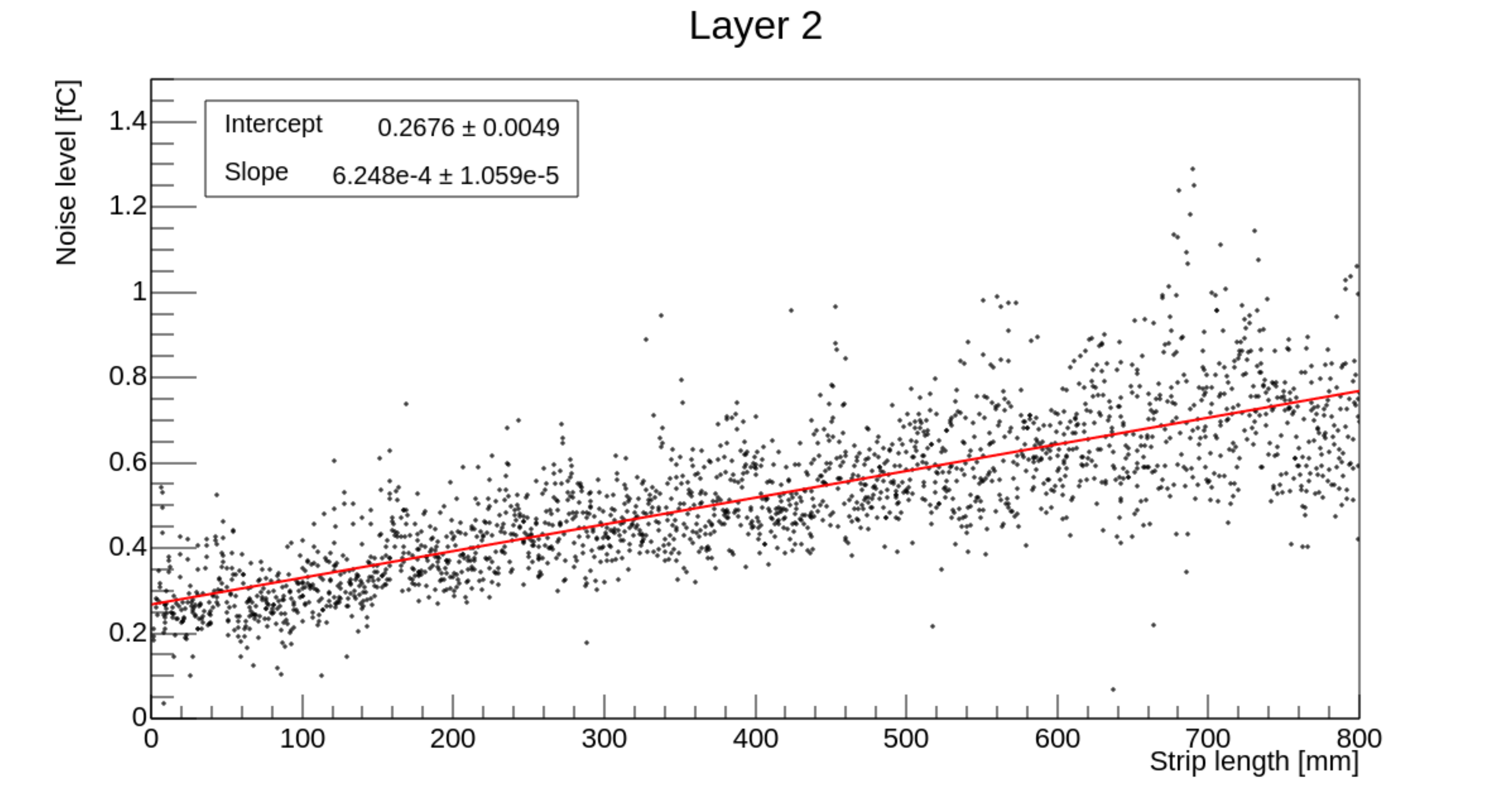}
  \caption{The trend of the noise vs the V strip length is due to the increasing capacitance of longer strips. The mechanical differences between the two layers determine the differences in slope and intercept. The parameters of the linear fit are reported in the legend.}
    \label{fig:noise_length}
\end{figure}
For instance, GUFI is used to measure the noise amplitude and noise spectrum. For the noise amplitude, a test pulse is sent to each TIGER channel, one by one, using the integrated test pulse generator. While sending the test pulse at a fixed rate, the threshold is swept over all the possible DAC values. Assuming that the noise amplitude is Gaussian, the rate curve can be fitted automatically to obtain an estimation of the standard deviation for the noise distribution.\\
With this procedure, the noise condition of the full test setup can be assessed (figure \ref{fig:noise_x}). The difference in noise levels of $\Upphi$ strips for layer 1 and layer 2 can be explained taking into account the different mechanical structure \cite{Balossino_2020}: layer 1 deploys a carbon fiber anode supporting structure, which increases the capacitive coupling of adjacent $\Upphi$ strips. Note that, as expected, the noise level of the V strips shows a trend following the strip length (figure \ref{fig:noise_length}).\\
To measure the noise spectrum, without saturating the Ethernet bandwidth, small sections of the setup can be acquired in trigger-less mode and the data analyzed via Fast Fourier Transformation in the frequency domain.\\
Optimization of the shielding and grounding scheme took advantage of the information regarding the noise and the communication stability of the system installed in Beijing. The same tools will be used during the final installation inside the BESIII spectrometer in order to optimize the setup grounding.\\
\begin{figure}
\centering
  \centering
  \includegraphics[width=0.9\linewidth]{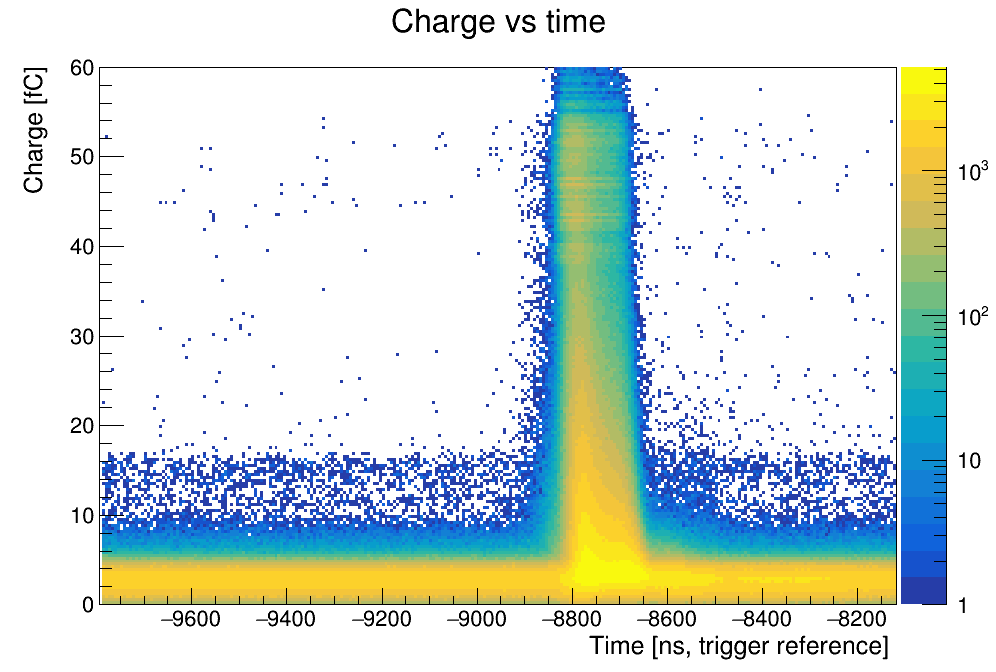}
  \caption{\changemarker{Hit charge versus hit arrival time with respect to the trigger arrival time on the layer 2, V strips. Data acquired during one week of cosmic rays acquisition. The broad central peak is the detector signal, while the low charge horizontal area is noise. }}
      \label{fig:status_plots_a}
\end{figure}
\begin{figure}
  \centering
  \includegraphics[width=0.9\linewidth]{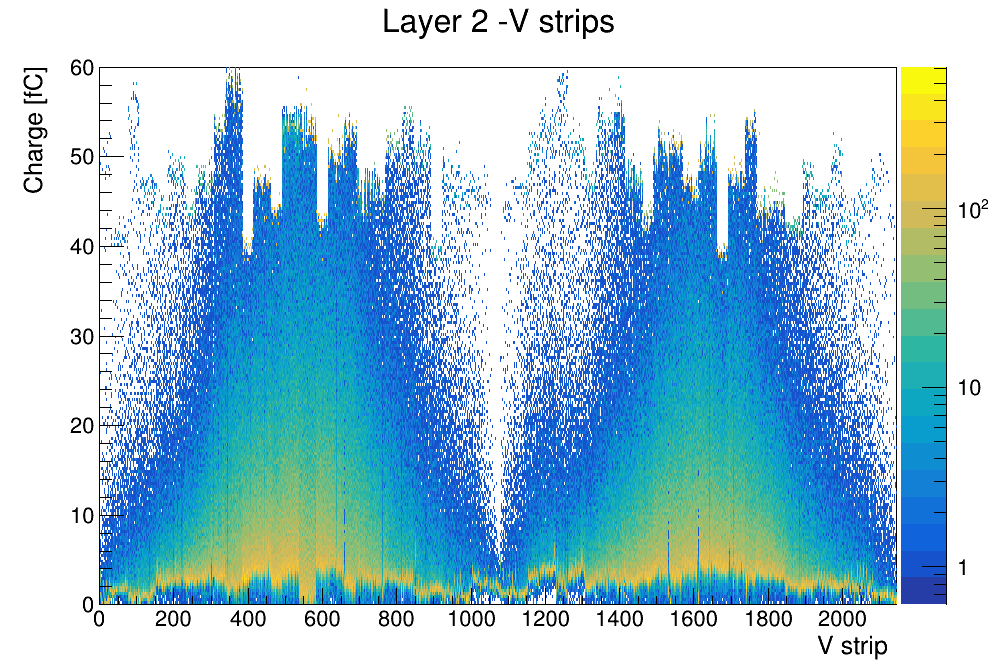}
  \caption{\changemarker{Charge on the layer 2 V strips in the signal time region. The charge is mostly distributed on the longest strips ( \SIrange[]{400}{700}{} and \SIrange[]{1500}{1800}{} ).}}
      \label{fig:status_plots_b}
\end{figure}
\changemarker{Important metrics for the detector status come from the data analysis performed at single-hit level by the DAQ software itself. Figure \ref{fig:status_plots_a} shows the charge of the hits versus the time with respect to the trigger, acquired during the cosmic ray data taking. With this kind of plot we can verify the detector signal, the good synchronization of the system and the absence of other structures in time beside the signal.
Figure \ref{fig:status_plots_b} shows the charge in the signal time region versus the strip position which permits to monitor the uniformity of the detector.\\}
\FloatBarrier
\section{Conclusions}
The CGEM-IT detector readout chain has been successfully developed. A dedicated on-detector and off-detector electronics has been designed according to the required performance. The TIGER ASIC reads directly the detector strips, providing the full digitized charge and time information. Its specification are tailored for the modern micropattern gas detectors. The GEMROC modules provide power, timing and control signals needed for TIGER operation and receive and process the output data of TIGER ASICs. Downstream the GEMROC modules, the VME-based GEM-DC  receive via optical links the trigger-matched data packets, assembling them into full events which are then made available to the VME-based BESIII DAQ system.\\
A dedicated system was devised for delivering and monitoring Low Voltage  and High Voltage power to the the readout electronics and to the detector. This system was designed to provide detailed control and monitoring of the operating voltages and currents.\\
All the components have already been tested and proved to achieve the required performance. The detector electronics is now in the last phases of its commissioning and optimization.\\
The whole readout system was designed focusing on great modularity and could be used for other innovative detectors.
\acknowledgments
We acknowledge the support of the European Commission in the RISE Project 645664-BESIIICGEM, RISE-MSCA-H2020-2014 and in the RISE Project  872901-FEST, H2020-MSCA-RISE-2019.\\
This paper is dedicated to the memory of our colleague Stefano Cerioni.
\newpage
\bibliography{biblio}

\providecommand{\href}[2]{#2}\begingroup\raggedright\begin{thebibliography}{10}

\bibitem{bepcII2009}
{\scshape BEPC II} collaboration, \emph{{BEPC} {II}: construction and
  commissioning},
  \href{https://doi.org/10.1088/1674-1137/33/s2/016}{\emph{Chinese Physics C}
  {\bfseries 33} (2009) 60}.

\bibitem{Ablikim2010}
{\scshape BESIII} collaboration, \emph{{Design and Construction of the BESIII
  Detector}}, \href{https://doi.org/10.1016/j.nima.2009.12.050}{\emph{Nucl.
  Instrum. Meth. A} {\bfseries 614} (2010) 345}
  [\href{https://arxiv.org/abs/0911.4960}{{\ttfamily 0911.4960}}].

\bibitem{White_paper}
{\scshape BESIII} collaboration, \emph{{Future Physics Programme of BESIII}},
  \href{https://doi.org/10.1088/1674-1137/44/4/040001}{\emph{Chinese Physics C}
  {\bfseries 44} (2020) 040001}.

\bibitem{Dong_2016}
M.-Y. Dong, Q.-L. Xiu, L.-H. Wu, Z.~Wu, Z.-H. Qin, P.~Shen et~al., \emph{Aging
  effect in the {BESIII} drift chamber},
  \href{https://doi.org/10.1088/1674-1137/40/1/016001}{\emph{Chinese Physics C}
  {\bfseries 40} (2016) 016001}.

\bibitem{Sauli2016}
F.~Sauli, \emph{{The gas electron multiplier ( GEM ): Operating principles and
  applications}},
  \href{https://doi.org/10.1016/j.nima.2015.07.060}{\emph{Nuclear Instruments
  and Methods in Physics Research, A} {\bfseries 805} (2016) 2}.

\bibitem{marcello2018}
S.~Marcello, M.~Alexeev, A.~Amoroso, R.~Baldini~Ferroli, M.~Bertani, D.~Bettoni
  et~al., \emph{{A new inner tracker based on GEM detectors for the BESIII
  experiment}},
  \href{https://doi.org/10.1142/S2010194518601199}{\emph{International Journal
  of Modern Physics: Conference Series} {\bfseries 48} (2018) 1860119}.

\bibitem{GEM_magnetic_field}
M.~Alexeev, A.~Amoroso, S.~Bagnasco, R.~B. Ferroli, I.~Balossino, G.~Bencivenni
  et~al., \emph{Triple {GEM} performance in magnetic field},
  \href{https://doi.org/10.1088/1748-0221/14/08/p08018}{\emph{Journal of
  Instrumentation} {\bfseries 14} (2019) P08018}.

\bibitem{Alexopoulos2010}
T.~Alexopoulos et~al., \emph{{Development of large size Micromegas detector for
  the upgrade of the ATLAS muon system}},
  \href{https://doi.org/10.1016/j.nima.2009.06.113}{\emph{Nucl. Instrum. Meth.
  A} {\bfseries 617} (2010) 161}.

\bibitem{Riccardo2016}
M.~Alexeev, A.~Amoroso, F.~Bianchi, M.~Bertani, D.~Bettoni, N.~Canale et~al.,
  \emph{{Development and test of a $\mu$TPC cluster reconstruction for a triple
  GEM detector in strong magnetic field}},  in \emph{2016 IEEE Nuclear Science
  Symposium, Medical Imaging Conference and Room-Temperature Semiconductor
  Detector Workshop (NSS/MIC/RTSD)}, pp.~1--4, 2016,
  \href{https://doi.org/10.1109/NSSMIC.2016.8069914}{DOI}.

\bibitem{design_2014}
\relax{BESIII Collaboration}, \emph{Conceptual Design Report - BESIII
  Cylindrical GEM Inner Tracker, version 1.0.1, Internal note}, 2014.

\bibitem{RIVETTI2019181}
A.~Rivetti, M.~Alexeev, R.~Bugalho, F.~Cossio, M.~D.~R. Rolo, A.~D. Francesco
  et~al., \emph{{TIGER}: A front-end {ASIC} for timing and energy measurements
  with radiation detectors},
  \href{https://doi.org/https://doi.org/10.1016/j.nima.2018.09.010}{\emph{Nuclear
  Instruments and Methods in Physics Research Section A: Accelerators,
  Spectrometers, Detectors and Associated Equipment} {\bfseries 924} (2019) 181
  }.

\bibitem{Geronimo}
G.~{De Geronimo}, J.~{Fried}, S.~{Li}, J.~{Metcalfe}, N.~{Nambiar}, E.~{Vernon}
  et~al., \emph{{VMM1—An} {ASIC} for micropattern detectors},
  \href{https://doi.org/10.1109/TNS.2013.2258683}{\emph{IEEE Transactions on
  Nuclear Science} {\bfseries 60} (2013) 2314}.

\bibitem{DIFRANCESCO2016194}
A.~D. Francesco, R.~Bugalho, L.~Oliveira, A.~Rivetti, M.~Rolo, J.~C. Silva
  et~al., \emph{\relax{TOFPET} 2: A high-performance circuit for \relax{PET}
  time-of-flight},
  \href{https://doi.org/https://doi.org/10.1016/j.nima.2015.11.036}{\emph{Nuclear
  Instruments and Methods in Physics Research Section A: Accelerators,
  Spectrometers, Detectors and Associated Equipment} {\bfseries 824} (2016) 194
  }.

\bibitem{fabio_tesi}
F.~Cossio, \emph{A mixed-signal ASIC for time and charge measurements with GEM
  detectors}, Ph.D. thesis, PoliTO, 2019.

\bibitem{ARRIVA}
Altera, \emph{Arria V GX, GT, SX, and ST Device Datasheet}, 12, 2013.

\bibitem{trigger}
Z.~{Liu}, W.~{Gong}, Y.~{Guo}, D.~{Jin}, {Lu Li}, Y.~{Lu} et~al., \emph{Trigger
  system of \relax{BESIII}},  in \emph{2007 15th IEEE-NPSS Real-Time
  Conference}, pp.~1--4, 2007.

\bibitem{DAQ_KLOE}
P.~Branchini, A.~Budano, A.~Balla, M.~Beretta, P.~Ciambrone, E.~D. Lucia
  et~al., \emph{Front-end {DAQ} strategy and implementation for the {KLOE}-2
  experiment},
  \href{https://doi.org/10.1088/1748-0221/8/04/t04004}{\emph{Journal of
  Instrumentation} {\bfseries 8} (2013) T04004}.

\bibitem{SY4527LC}
CAEN, \emph{SY4527LC, Universal Multichannel Power Supply System, Datasheet}.

\bibitem{A2517}
CAEN, \emph{A2517, 8 Channel 5 V/15 A (50 W) Individual Floating Channel
  Boards, Datasheet}.

\bibitem{SY5527}
CAEN, \emph{SY5527, Universal Multichannel Power Supply System, Datasheet}.

\bibitem{A2519}
CAEN, \emph{A2519, 8 Channel 15V/5 A (50 W) Individual Floating Channel Boards,
  Datasheet}.

\bibitem{A1515}
CAEN, \emph{A1515, 16/14 Channel 1-1.3kV (1-3 mA) Individual Floating Channel
  Dual Range Boards for Quadruple and Triple GEM detectors, Datasheet}.

\bibitem{Influx}
InfluxData Inc., \emph{Influx DataBase}.

\bibitem{Grafana}
Grafana Labs, \emph{Grafana}.

\bibitem{Farinelli_2020}
R.~Farinelli, M.~Alexeev, A.~Amoroso, S.~Bagnasco, R.~B. Ferroli, I.~Balossino
  et~al., \emph{{GRAAL}: Gem reconstruction and analysis library},
  \href{https://doi.org/10.1088/1742-6596/1525/1/012116}{\emph{Journal of
  Physics: Conference Series} {\bfseries 1525} (2020) 012116}.

\bibitem{Balossino_2020}
I.~Balossino, \emph{Investigation and improvements of the mechanical structure
  of cylindrical {GEMs} for the {BESIII} experiment},
  \href{https://doi.org/10.1088/1748-0221/15/08/c08013}{\emph{Journal of
  Instrumentation} {\bfseries 15} (2020) C08013}.

\end{thebibliography}\endgroup
\bibliographystyle{JHEP}
\end{document}